# 3D Reconstruction of Bias Effects on Porosity, Alignment and Mesoscale Structure in Electrospun Tubular Polycaprolactone

Y. Liu[a], F.J. Chaparro[b], Z. Gray[b], J. Gaumer[c], D. B. Cybyk[d], L. Ross[e], P. Gosser[a], Z. Tian[a], Y. Jia[a], T. Dull[a], A.L. Yarin[f], J. J. Lannutti[a,d,g]

[a] The Ohio State University, Department of Materials Science and Engineering, 116 W 19th Avenue, Columbus, OH 43210, USA
[b] Nanoscience Instruments, 10008 S. 51st Street, Suite 110, Phoenix, AZ 85044, USA
[c] Tosoh SMD, Inc. 3600 Gantz Road Grove City, OH 43123, USA
[d] The Ohio State University, Department of Biomedical Engineering, 116 W 19th Avenue, Columbus, OH 43210, USA
[e] Columbus Academy, 4300 Cherry Bottom Road, Gahanna, OH, 43230, USA
[f] University of Illinois Chicago, Department of Mechanical and Industrial Engineering, 842 West Taylor Street, MC 251, Chicago, IL 60607, USA
[g] Center for Chronic Brain Injury Program, The Ohio State University, Columbus, OH 43210, USA

## Abstract

Porosity variations in tubular scaffolds are critical to reproducible, sophisticated applications of electrospun fibers in biomedicine. Established laser micrometry techniques produced ~14,000 datapoints enabling thickness and porosity plots versus both the azimuthal ($\Phi$) and axial (Z) directions following cylindrical mandrel deposition. These 3D datasets could then be 'unrolled' into 'maps' revealing variations in thickness and porosity versus 0, -5, and -15 kV collector bias. As bias increases, thinner, more 'focused' depositions occur. At 0 kV bias, maximum thickness coincides with maximum porosity; at -5 kV bias, maximum thickness coincides with minimum porosity. Porosity maps show that at 0 kV, a concave-down central region of higher (~93-94%) porosity exists bounded on either side by roughly symmetric, parabolic decreases to ~87-89% within ~15v%. At -5 kV, a different concave-up character occurs, showing a central porosity of ~82-84% bounded by symmetric, parabolic increases in porosity to ~85-86%. At -15 kV, the porosity profile shows either concave-up or linear behavior. Simultaneous decreases in net porosity versus bias (91.1%@0kV > 83.4%@-5kV > 80.2%@-15 kV) are sensible, but significant changes in the distribution were unexpected. Surprisingly, at 0 kV, extensive mesoscale surface roughness is evident. Optical profilometry revealed unique features ~1600×420 μm in size, standing ~210 μm above the surrounding surface. These shrink to only ~440×150 μm in size and ~30 μm higher at -5 kV bias and disappear entirely at -15 kV. Scanning electron microscopy (SEM) resolved these into novel, localized 'domains' containing tightly aligned fibers oriented parallel to the mandrel axis. Observation of 'curly' fibers in the SEMs following -5 and -15 kV indicate buckling instabilities. This agrees with prior observations of residual solvent effects: increased bias causes faster motion toward the mandrel, meaning (1) its solvent content upon arrival is higher, leading to lower viscosities less resistant to buckling/compaction, (2) higher velocities during deposition cause both decreased porosity/"denser packing" and increased buckling. Unexpectedly, we also observed substantial orientation along the mandrel axis. By modifying classical bending instability models to incorporate cylindrical electric fields, simulation revealed that horizontal components in the modified electric field alter bending loop shape, causing the observed alignment. This provides a new, easily utilized tool enabling facile, efficient tuning of orientation.

**Introduction**

An applied bias is necessary to deposit electrospun fiber produced at the modified Taylor Cone toward a specific collection surface [1-4]. Bias values are significant to tissue engineering because they exert control over porosity, a factor directly relevant to cell penetration, a well-recognized challenge limiting the utility of electrospun scaffolds [5-9]. While a wealth of knowledge has been generated at the small observable via SEM, quantification of larger-scale 3D porosity or orientation variations versus collector bias has not yet appeared in the literature. In addition, microstructure and porosity resulting from electrospinning connect as the amount of porosity is generally inversely proportional to the number of fibers occupying a given volume. However, relationships between fiber diameter, polymer moduli, interfiber friction and overall porosity are far from satisfactorily explored. In the context of tissue engineering, a broad range of such bias values have been utilized: 5-25 [10, 11], 12 [12], 17.5 [13], 20 [14], and 21 kV [15]. An absence of understanding of the associated effects of these widely varying levels of bias on porosity is likely a factor preventing successful, uniform cellular infiltration and remodeling. In addition, the efficiency of fiber deposition on the collector surface during electrospinning increases with applied bias, creating a significant driving force toward higher voltages, especially if costly polymeric or biological compounds are involved.

The geometry of collectors used to produce these deposits vary widely but rotating cylindrical drums and mandrels enable specific fabrication of scaffolds for vascular, cardiovascular, urological, neural, and tracheal applications [16-20]. For any scaffold, local values of porosity directly control cell infiltration, mechanical properties, responses to cyclically applied stresses, degradation uniformity, and net resorption time *in vivo*.

We recently developed a novel technique that overcomes the limitations of traditional porosimetry methods to link microstructure and specific conditional influences during scaffold fabrication [21]. High solvent contents present in arriving fiber during deposition decreased porosity by increasing fiber collapse or densification post-spinning. While PCL-only-based solutions were relatively insensitive to humidity, the addition of Rose Bengal suppressed solvent evaporation under higher levels of humidity due to its net hydrophilicity, rendering porosity exquisitely sensitive to higher humidities [21]. Rose Bengal also enabled tracking to prove the existence of asymmetric deposition during two-needle electrospinning. In the context of more complex, multi-layered [22] shapes associated with tissue engineering, uniform deposition onto three-dimensional mandrels can easily be hindered by variable process conditions [23, 24].

We sought to utilize this technique for porosity determination establishing the connection between applied bias and the resulting microstructures in both the axial and azimuthal directions during the production of tubular scaffolds. A constant 10 kV bias was applied to the needle; 0, -5 or -15 kV biases were applied to the mandrel. The needle to mandrel distance was held constant at 20 cm. As before [21], laser micrometry detailed mandrel surface geometry (1) before deposition, (2) the fiber geometry spun onto that mandrel after deposition, and – finally – (3) the film geometry produced following complete fiber sintering. These steps create descriptions of variations in thickness and porosity across the entire deposit as a function of applied bias. Optical profilometry (OP) was used as an intermediate characterization technique to identify features of interest; scanning electron microscopy (SEM) was then used to examine these features in greater detail to establish finer aspects of microstructure and fiber alignment. Interestingly, we observed a novel feature present immediately after electrospinning: small, localized regions consisting of highly aligned fiber bundles.

In examining microstructure at a larger scale, we discovered that statistically, predominant fiber orientations were perpendicular to the direction of rotation. Given the relatively broad use of cylindrical collectors in electrospinning, it is surprising that few, if any, modeling investigations have focused on tubular configurations, particularly in regards to fiber alignment. Bending instabilities were modeled using a Lagrangian discretization of the fiber jet into a series of viscoelastically connected nodes to produce a system of equations representing fiber motion toward an infinitely long, uniformly charged cylinder. The electric field **E** around the cylindrical collector is then shown to contain a non-zero component in the horizontal plane, facilitating the experimentally observed alignment of deposited fibers parallel to the mandrel axis. With bead insertion and adaptive refinement procedures adapted from Lauricella et al. [25-27], it was revealed that the horizontal cross-section of the bending cone evolved from a circular to an elliptical shape under the influence of this non-uniform field **E**. During deposition, this elliptical character resulted in a preferred orientation aligning along the mandrel axis.

**Materials and Methods**

*Solution Preparation*

A 5 wt% PCL (Sigma-Aldrich, St. Louis, MO, $M_n$ = 80,000) was dissolved in 1,1,1,3,3,3-hexafluoro-2-propanol (HFP) (Oakwood Chemicals, West Columbia, SC) while mechanically stirring at 300 rpm and room temperature (~20°C) for 8 h, producing a homogeneous solution inside a 250 mL Erlenmeyer flask. Once the solid was completely dissolved and the solution appeared well mixed, it was transferred to a 60 cm$^3$ plastic syringe (BD Luer-Lok, Franklin Lakes, NJ).

*Solution Characterization*

The solution was characterized with regard to viscosity, surface tension and solids content. Viscosity was measured using a VisoQC™ 300 Type L (Anton-Paar, Ashland, VA) equipped with a low volume cup using a CC12 spindle and measured at 23°C (n = 3). Surface tension was measured at room temperature using a Theta Lite optical tensiometer (Biolin Scientific, Gothenburg, Sweden). The analysis was performed using a 22-gauge needle using a pendant drop of the solution ($n = 3$). The surface tension was recorded at 1.7 frames per second for at least 10 seconds using OneAttension software (v4.0.5). Solid content was determined using a moisture analyzer (MB27, Ohaus, Parsippany, NJ) at a temperature of 58°C with a minimum of a 3 g sample ($n = 3$). These specific results are found in Table S2.

*Preparation of Electrospun Scaffolds*

The as-prepared PCL solution was electrospun within a Fluidnatek® LE-100 unit (Bioinicia S.L., Valencia, Spain) equipped with an environmental control unit enabling precise control over temperature (T), relative humidity (RH), air flow rate, and applied voltage (-30 to 30 kV). A 316L mandrel 0.95 cm in outside diameter and 30 cm in length (McMaster-Carr, Catalog# 89325K93) was used as the collection substrate. Electrospinning conditions were as follows: the PCL solution was transferred from the syringe to a 20-gauge blunt needle (0.61 ID x 12.7 mm in length; EFD, East Providence, RI) using a 1.6 mm outside diameter polytetrafluoroethylene (PTFE) tubing at a flow rate of 6 mL h$^{-1}$ and an applied +10 kV bias. Three independent voltage conditions were applied to the collector: 0 (grounded), -5 and -15 kV. The chamber was maintained 25°C and 30% RH. To properly remove evaporated solvent and avoid accumulation inside the chamber, an airflow of 80 m$^3$ h$^{-1}$ was used. Vertical spinning was used; the needle was held at a 20 cm distance from the collector oriented at 45° [21]. For all three biases, the mandrel was rotated clockwise at 200 rpm. Once electrospinning was initiated, the fibers were collected in the central area of the mandrel under static conditions for 15 minutes ($n = 4$; 3 for laser profiling/1 for microstructural analysis).

*Electrospun Scaffold Densification*

To enable porosity determination, as-spun samples were sintered as before [21] to full density by exposure to 65°C for 3 hours. Prior work with electrospun PCL sintering showed us that the initial electrospun porosity is rapidly eliminated by sintering below the melting point (60°C) [28-30].

*Laser Profiling*

The profiling of sample thickness and porosity followed the same principle as previously reported [21] but expanded to include the azimuthal/circumferential direction. The mandrel was first loaded onto a custom-built profiling system consisting of a vertical linear actuator (ALS20020, Aerotech), a rotary stage (Model 20501, Parker Automations), and a laser micrometer (TLA122s, Laserlinc). The rotary stage presents the mandrel to the micrometer at 72 different angles, each 5 degrees apart providing a full 360-degree interrogation. Following each rotation, the actuator translates the micrometer 200 mm along the mandrel to perform a scan containing roughly ~201 datapoints. Each mandrel was scanned three times: as a bare rod, following deposition and after PCL sintering to full density [28].

Raw output from the profiling system was first transformed into a point cloud containing ~14,000 points within a cylindrical coordinate system. Here, the positional feedback from the rotary stage and linear actuator corresponds to azimuthal and axial coordinates, respectively, and diametric measurements are halved to form the radial coordinate assuming two-fold rotational symmetry. The point cloud so generated is then used to construct a continuous surface using linear interpolation to enable the subtraction of the bare mandrel profile (R) from as-deposit and densified profiles. This operation yields the as-deposited (H) and densified (h) thicknesses, respectively. From here, the porosity profile can be calculated using equation (*1*):

$$p = \frac{V_{As-deposit} - V_{Densified}}{V_{As-deposit}} \times 100\% = \left[1 - \frac{h(2R + h)}{H(2R + H)}\right] \times 100\% \quad (1)$$

All thickness and porosity profiles are evaluated along a 72-by-201 evenly spaced matrix that emulates the originally measured locations. Any porosity values where the as deposited thickness, densified thickness, or differences between the two is less than or equal to twice of the accuracy of the laser micrometer (±2.5 μm) were discarded.

The mean roughness ($R_a$) was estimated along the circumferential/azimuthal direction using equation (*2*) [31]:

$$Ra_i = \frac{1}{n}\sum_{j=1}^{n}|T_{i,j} - \bar{T}_i| \quad (2)$$

Where $T_{i,j}$ is the *i*th thickness along the axial direction at the *j*th rotation, $\bar{T}_i$ is the *i*th mean thickness across all rotations, and n=72. Then the mean of all $Ra_i$ is reported as the average Ra of a sample.

*Optical Profilometry*

Samples either still in place on the rod or removed from the rod and adhered to a SEM stub were positioned on the stage of the optical profiler (OP) (ZETA-20, KLA, Milpitas, CA). Scans of the 0, -5, and -15 kV samples were obtained at both 5X and 20X magnifications. These areas were then marked in such a way that OP and SEM analyses could be obtained from the same location.

*Scanning Electron Microscopy*

Samples were placed on conductive carbon tape (TED Pella, Redding, CA) adhered to aluminum SEM sample mounts (TED Pella, Redding, CA) and sputter coated for 30 seconds with gold at an emission current of 25 mA. Sputtering was then repeated two additional times at 30 second intervals to avoid microstructural damage. Microstructure was then observed under an SEM (Thermo Scientific Phenom XL Desktop SEM) at accelerating voltages of either 10 or 15 kV and at magnifications between 500X and 50,000X. Fiber diameter was analyzed using a Fibermetric software package (v2.3.4.0) that used > 2,500 readings taken from four images at magnifications of 3,000 X. Stitched images were developed by scanning total areas of ~1,500 x 3,800 µm, ~800 x 3,000 µm, and ~350 x 3,000 µm for the 0, -5 and -15 kV biased depositions, respectively. Fifty individual images were taken at a magnification of 640 X (0 kV) or 830 X (-5 kV and -15 kV) and stitched together using the Image Compositor Editor software (v2.0.3.0) using a structured panorama with camera motion set to 'planar motion with skew' to maintain this view of the final structure. To minimize imperfections in the composite image, the 'auto overlap' option was selected that lined up all 50 tiles within a continuous stitched image. Fiber orientation was measured from the stitched images using the ImageJ (1.52p) plug-in OrientationJ

(v2.0.5) [32]. Fiber bundles were imaged by placing the sample on a low profile 45° SEM mount (#16104, TED Pella, Redding, CA) prior to coating using the previous parameters.

*Modeling of Bending Instability*

Bending instabilities during electrospinning are modeled using a discretized system based on quasi-one-dimensional equations of free liquid jets. Lagrangian coordinates are used and, accordingly, a polymer jet is presented as a series of viscoelastically connected nodes [2, 25, 27, 33]. For each node, the equations of motion (*3-5*) can be written as:

$$\frac{\partial \boldsymbol{r}_i}{\partial t} = \boldsymbol{v}_i \tag{3}$$

$$m_i \frac{\partial \boldsymbol{v}_i}{\partial t} = \pi a_{i+1,i}^2 \sigma_{i+1,i} \hat{\boldsymbol{r}}_{i+1,i} + \pi a_{i-1,i}^2 \sigma_{i-1,i} \hat{\boldsymbol{r}}_{i-1,i} + \frac{\pi \alpha (a_{i+1,i} + a_{i-1,i})^2}{4|\boldsymbol{k}_i|} \hat{\boldsymbol{k}}_i + e_i \boldsymbol{E} + \sum_{j, j \neq i} \frac{e_i e_j}{|\boldsymbol{r}_{i,j}|^2} \hat{\boldsymbol{r}}_{i,j} \tag{4}$$

$$\frac{\partial \sigma_{i,j}}{\partial t} = \frac{G}{|\boldsymbol{r}_{i,j}|} \frac{\partial |\boldsymbol{r}_{i,j}|}{\partial t} - \frac{G}{\mu} \sigma_{i,j} \quad (5)$$

where $\boldsymbol{r}_i$, $\boldsymbol{v}_i$, $m_i$, and $e_i$ are the position, velocity, mass, and electric charge of the $i^{th}$ node; $\alpha$, $G$, $\mu$, and $\boldsymbol{E}$ are the surface tension coefficient, elastic modulus, viscosity, and the external electric field strength, respectively. In addition, $\boldsymbol{r}_{i,j} = \boldsymbol{r}_i - \boldsymbol{r}_j$ is the vector from the $j^{th}$ to the $i^{th}$ node; $a_{i,j}$ and $\sigma_{i,j}$ are the radius and stress associated with jet segment $\boldsymbol{r}_{i,j}$ when $|i - j| = 1$; $\boldsymbol{k}_i$ is the vector from the $i^{th}$ node to the local curvature center of the jet axis; the "hat" circumflex denotes the unit vector of an arbitrary non-zero vector, e.g., $\hat{\boldsymbol{u}} = \boldsymbol{u}/|\boldsymbol{u}|$.

The collector was modeled as a charged rod of infinite length. Correspondingly, the external field strength $\boldsymbol{E}$ is written as a function of the applied bias (difference) $\varphi_0$, the needle to mandrel (axis) distance $h$ and the mandrel radius $a_c$:

$$\mathbf{E} = -\frac{\varphi_0}{\ln(h/a_c)} \frac{1}{|r_{yz}|} \hat{r}_{yz} \qquad (6)$$

where $\mathbf{r}_{yz} = r_y \hat{y} + r_z \hat{z}$ is the projection of *r* on the *yOz* plane.

**Results**

Figure 1 shows the thickness profiles generated from ~14,000 datapoints (per sample) established using laser micrometry for the selected depositions, plotted against both the azimuthal (Φ) and axial (Z) directions of the mandrel. The use of the cylindrical coordinate system allows for 'unrolling' to illustrate the behavior of the thickness data. At 0 kV, extensive surface roughness in the middle of deposition (approximately 50<Z<110 mm) is obvious. This level of roughness is absent from both the -5 and -15 kV depositions.

By manipulating the thickness data of all samples as described in the *Materials and Methods* section, the roughness values $R_a$ for all three biases were established and are reported in Table 1. This Table shows that the roughness of the 0 kV deposition is distinctly significantly ($p<0.001$) and 5-10 times rougher than the other two (statistically indistinguishable, $p>0.05$) biases. The reported values are averages taken across the entire deposition; at 0 kV, local values of $R_a$ in the central region (approximately $50 < Z < 110$ mm) exceed 20 μm (data not shown). Figure 2 provides visual confirmation of both the decreasing values of roughness and the level of focused deposition (inversely proportional to FWHM in Table 1) versus the net bias.

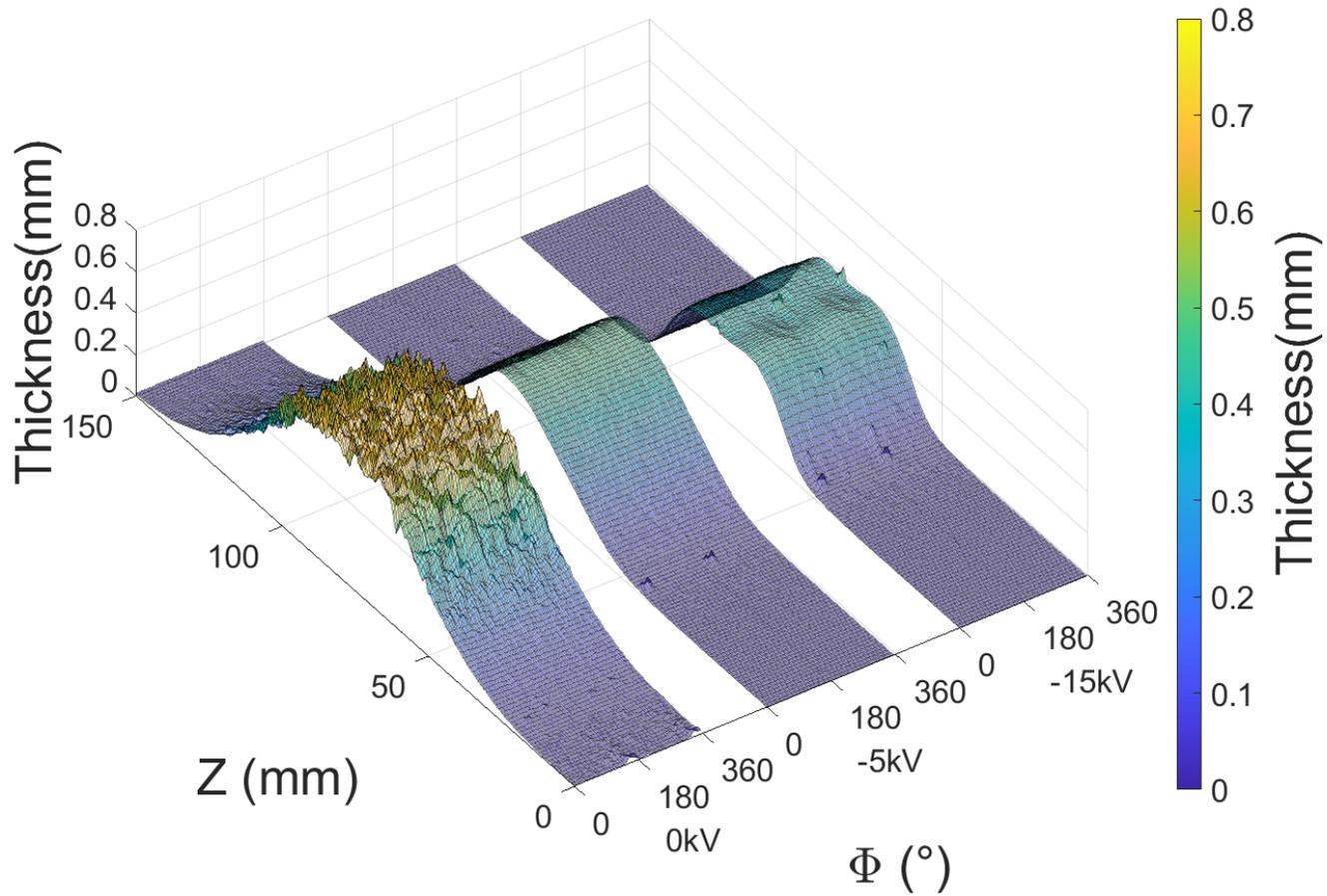

Figure 1. Three-dimensional representation of thickness variations for the 0, -5 and -15 kV depositions. The height and width of these depositions clearly decrease as collector bias decreases.

Table 1. Deposit thickness, FWHM, average porosity and fiber diameter versus applied bias. Values in the same column with different capital letters (A, B, C), lower case letters (a, b, c) or Greek letters (α, β, γ) in superscripts are statistically different at the 0.001, 0.01 or 0.05 significance levels, respectively.

|  | Deposit Peak Thickness (mm)* | Deposit FWHM (mm)* | Average Porosity (%)* | $R_a$ (μm)* | Fiber Diameter (nm)** |
|---|---|---|---|---|---|
| 0 kV | 0.7579±0.0323$^A$ | 56.3±4.7$^{aα}$ | 91.1±1.4$^A$ | 9.8±0.7$^A$ | 1862[1453-2419]$^{Bβ}$ |
| -5 kV | 0.4293±0.0426$^{Bα}$ | 46.9±3.9$^β$ | 83.4±0.6$^{Bα}$ | 1.4±0.4$^{Bα}$ | 1908[1542-2398]$^{Bα}$ |
| -15 kV | 0.4409±0.0200$^{Bα}$ | 37.0±1.7$^{bγ}$ | 80.2±0.8$^{Bβ}$ | 1.3±0.4$^{Bα}$ | 2187[1803-2605]$^A$ |
| p-value | 2.790*10$^{-5}$ | 2.016*10$^{-3}$ | 2.813*10$^{-5}$ | 1.306*10$^{-6}$ | 1.204*10$^{-59}$ |

*Mean ± standard deviation of 3 individual samples; statistical difference analyzed via one-way ANOVA with Tukey's multiple comparison test.
**Median [interquartile range] of ~2,500 individual measurements; statistical difference analyzed via Kruskal-Wallis test followed by Dunn's test with Benjamini-Hochberg correction for pairwise comparison.

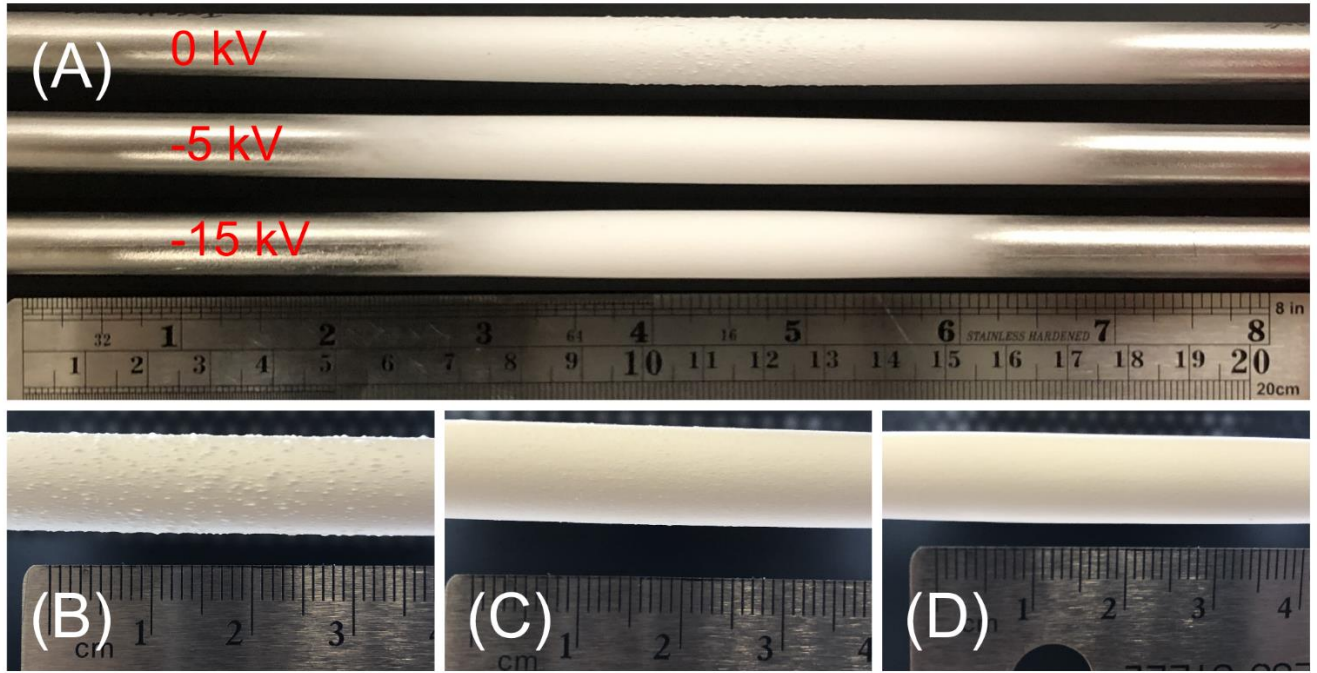

Figure 2. Optical images enabling side-by-side comparison of different collector biases (A), as well as higher magnifications of the 0 (B), -5 (C) and -15 (D) kV bias depositions. The 0 kV deposition is considerably wider, higher and rougher than the other two, consistent with the observations in Figure 1. Higher net bias results in more focused depositions and the visual elimination of the roughness.

While there is variance in thickness (perceived as roughness) in the azimuthal /rotational directions– especially in the 0 kV bias depositions – we could not identify statistically significant correlations between the two. This is anticipated as the spinning mandrel should effectively standardize (in a statistical sense) azimuthal thickness distributions during deposition. In Figure 3, the 3D data (Figure 1) for all samples is averaged and plotted against the axial direction, with the origin chosen to represent the peak of as-deposited thickness; outliers that are more than three median absolute deviations (MAD) [34] away were removed. As is evident, both the 0 and -5 kV thickness profiles closely follow a gaussian function ($R^2>0.97$) where good consistency was achieved within each bias applied in the Fluidnatek LE-100 electrospinning unit. For the -15 kV deposition, a Gaussian fit was still appropriate ($R^2>0.92$), although two of the depositions exhibited minor bimodal character.

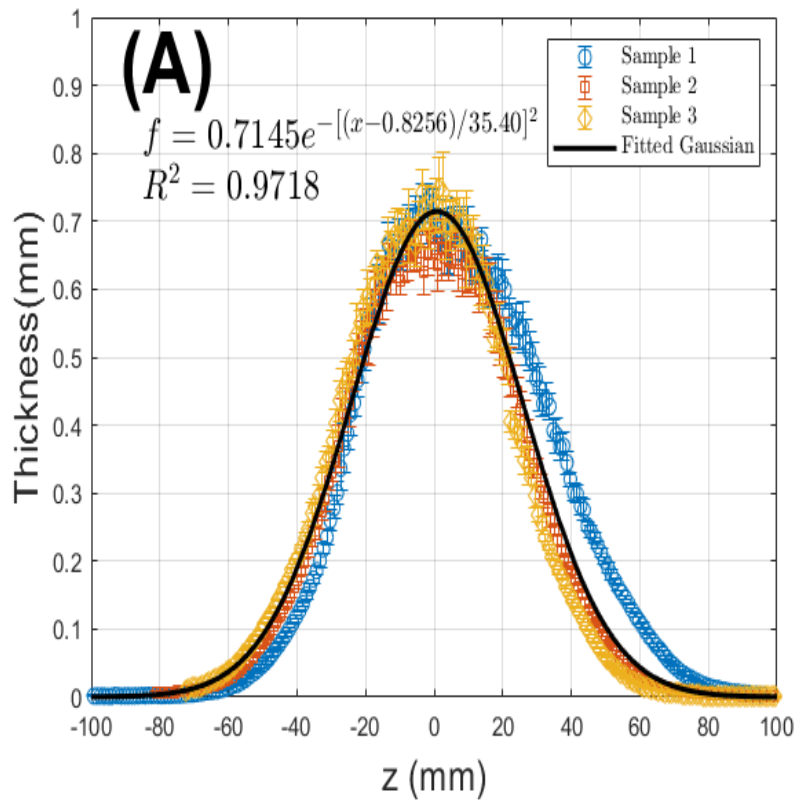

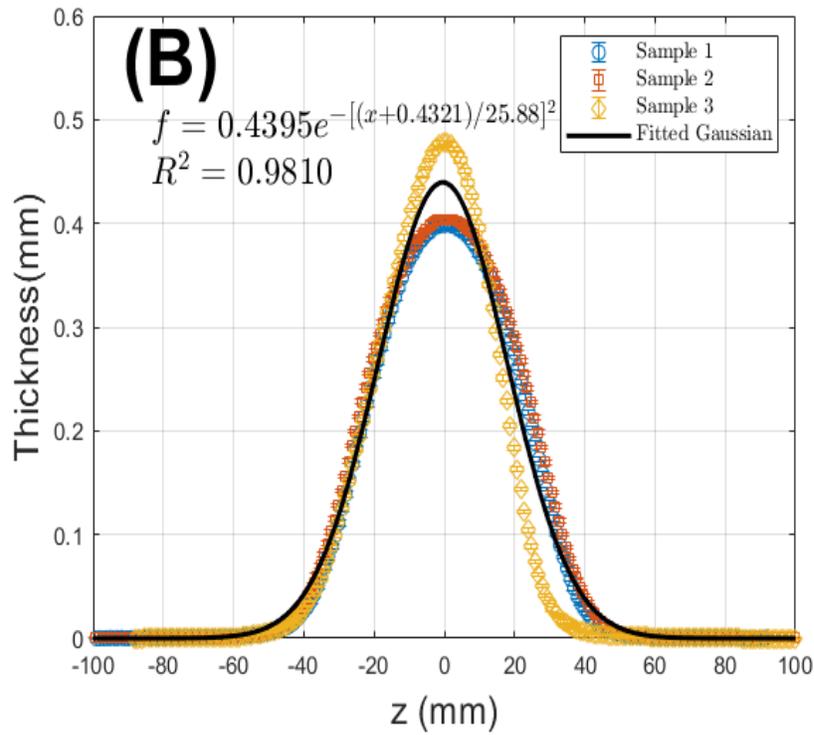

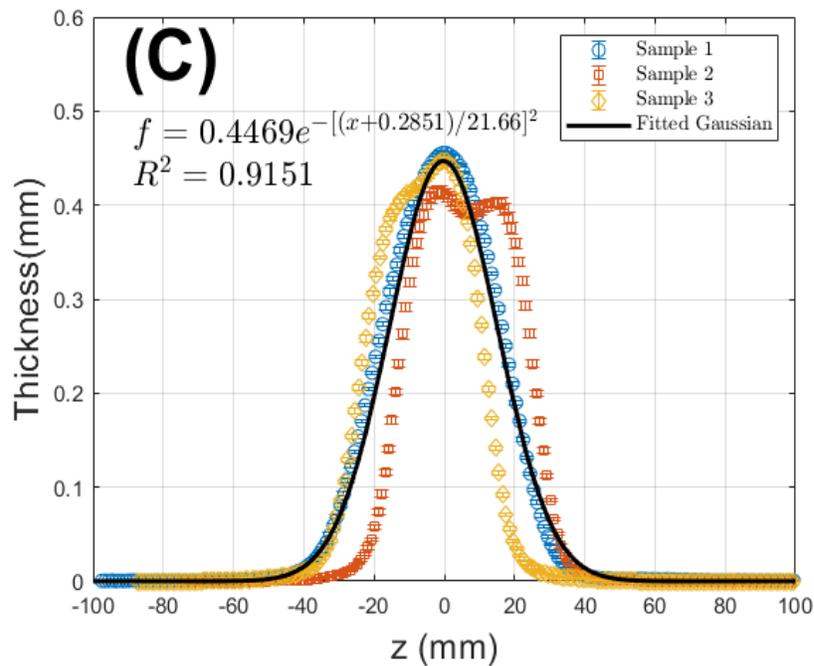

Figure 3. As-deposited thickness profiles of three (3) individual samples electrospun under 0 (A), -5 (B), and -15 (C) kV collector biases. To highlight deposition symmetry, the x-axis represents the axial direction of the mandrel, and the origin (z=0) is the point where peak thickness values were recorded. Each datapoint and error bar represents mean ± one standard deviation of 72 thickness values (minus outliers) at different azimuth/rotations for the corresponding axial position (z). In (A), the increase in error bar size scales with the roughness observed in Figures 1 and 2.

In Figure 3, a bias of 0 kV produces the thickest (~0.7 mm) peak deposition value (Figure 3A). The roughness trends shown in Figures 1 and 2A are also present in this data; the increase in error bar sizes begins and ends at approximately the same position on each mandrel (-20<z<30 mm). At -5 kV bias, peak thickness decreases substantially (to ~0.45 mm) relative to the 0 kV deposition value (Figure 3B). The optical image (Figure 2C) at this bias shows a visually smoother surface than that produced at 0 kV (Figure 2B). Finally, the -15 kV bias displays approximately the same peak thickness (~0.45 mm) as the -5 kV but shows slightly more variability (Figure 3C); Figure 2D also shows a surface visually smoother than that produced at 0 and -5 kV. The values of full-width-at-half-maximum (FWHM) (Table 1) confirm visual observations (Figure 2A) show that the depositions become more 'focused' and occupy smaller areas as the collector bias decreases.

The peak thickness and FWHM summarized in Table 1 show significant ($p<0.001$ for thickness, $p<0.01$ for FWHM) differences across the applied biases. For pairwise comparison, the difference in peak thickness between 0 and either -5 or -15 kV is not only distinctly significant ($p<0.001$) but nearly twice that of the other two biases (758 vs. 429 or 440 μm) that is not statistically distinguishable ($p>0.05$). For FWHM, the difference between 0 and -15 kV is significant ($p<0.01$), while either of the two groups is only moderately ($p<0.05$) significantly different from -5 kV. The mean of FWHMs decreases smoothly with bias: 0 kV/56.3 mm, -5 kV/46.9 mm, and -15 kV/37.0 mm.

The median and interquartile ranges of fiber diameter are also reported in Table 1 and follow the ranked trend of 0 (1862 [1453-2419] nm) < -5 (1908 [1542-2398] nm) < -15

(2187[1803-2605] nm) kV. As the distributions of fiber diameter are highly skewed (Figure S1), the Kruskal–Wallis test – a nonparametric equivalent of standard one-way ANOVA – was used to show a distinctly significant difference (p<0.001) versus collector bias. A pairwise comparison shows that the difference between -15 kV and the other two biases is highly significant (p<0.001); the difference between 0 and -5 kV is only slightly significant (p<0.05).

In Figures 4A and 4B, small regions we found to consist of localized domains of highly oriented fiber running parallel to the mandrel axis could be observed following the 0 kV depositions. Outside of these regions, individual fibers splay outwards. Figure 4C reveals a higher magnification of this region in the tilted view, showing that the center of each region consists of a substantial number of individual fibers stacked parallel to each other and to the axial direction of the mandrel. In contrast, at -5 kV bias, Figures 4D and 4E show that these domains are still present but appear to be much smaller and have a much more limited vertical component. To quantify these differences, we conducted optical profilometry (Figure 5) to create a mesoscale view of the surface of the as-spun depositions showing that these features are typically ~1600×420 μm in size and ~210 μm higher than the surrounding surface following the 0 kV deposition. At -5 kV, these features are still present but become much less frequent and shrink to only ~440×150 μm in size and ~30 μm higher than the surrounding surface. At -15 kV, these features are absent in both the SEM and OP images. Less noticeable is that the degree of fiber 'curviness' increases substantially from 0 to -15 kV; the fibers at 0 kV generally move in roughly straight lines on the scale of the SEMs. At -15 kV, nearly all visible fibers are curved or 'curly,' in some cases even on the scale of a few microns. The -5 kV deposition shows an intermediate morphology between the two extremes.

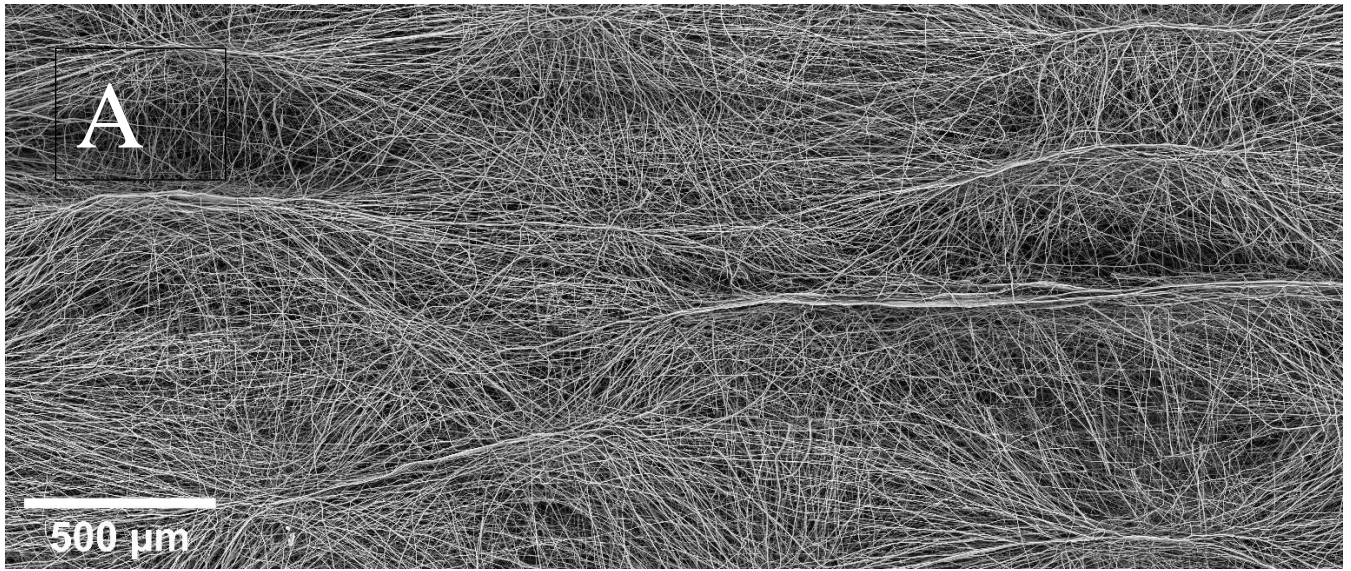

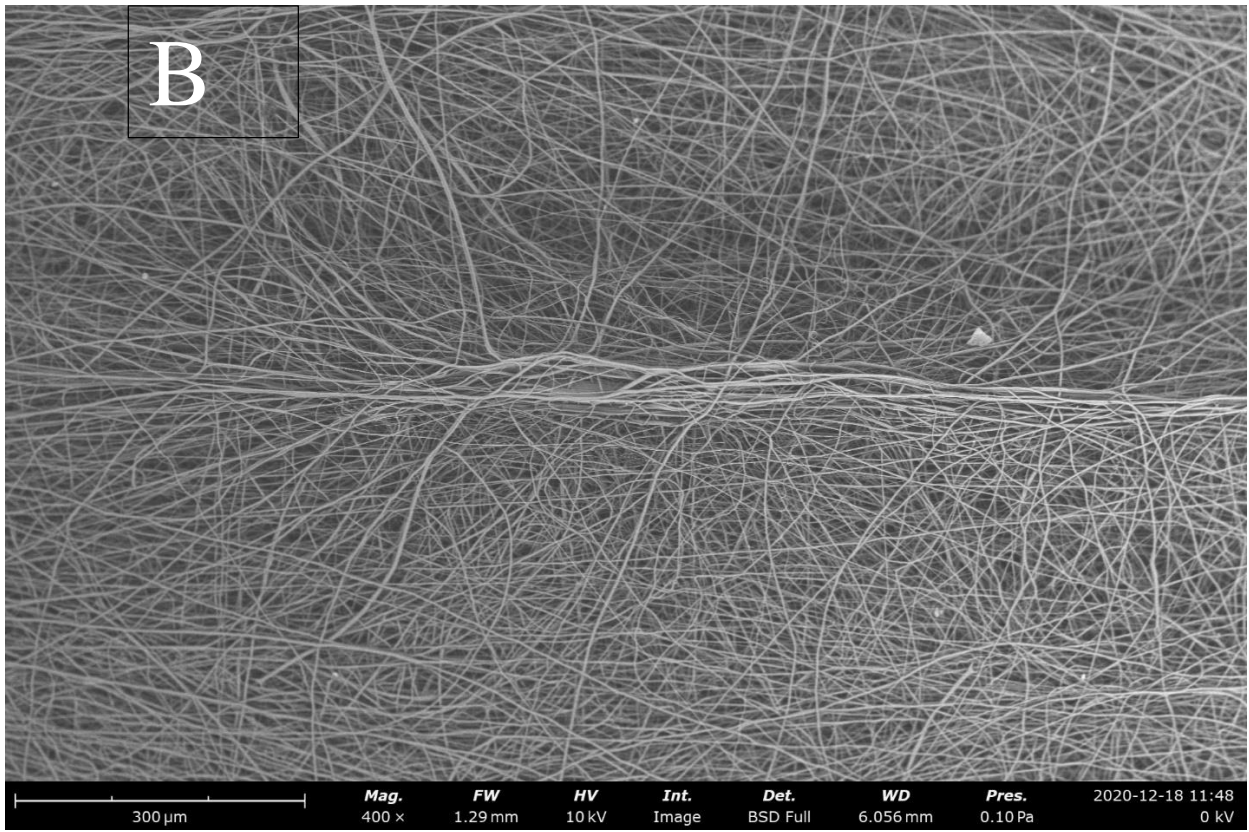

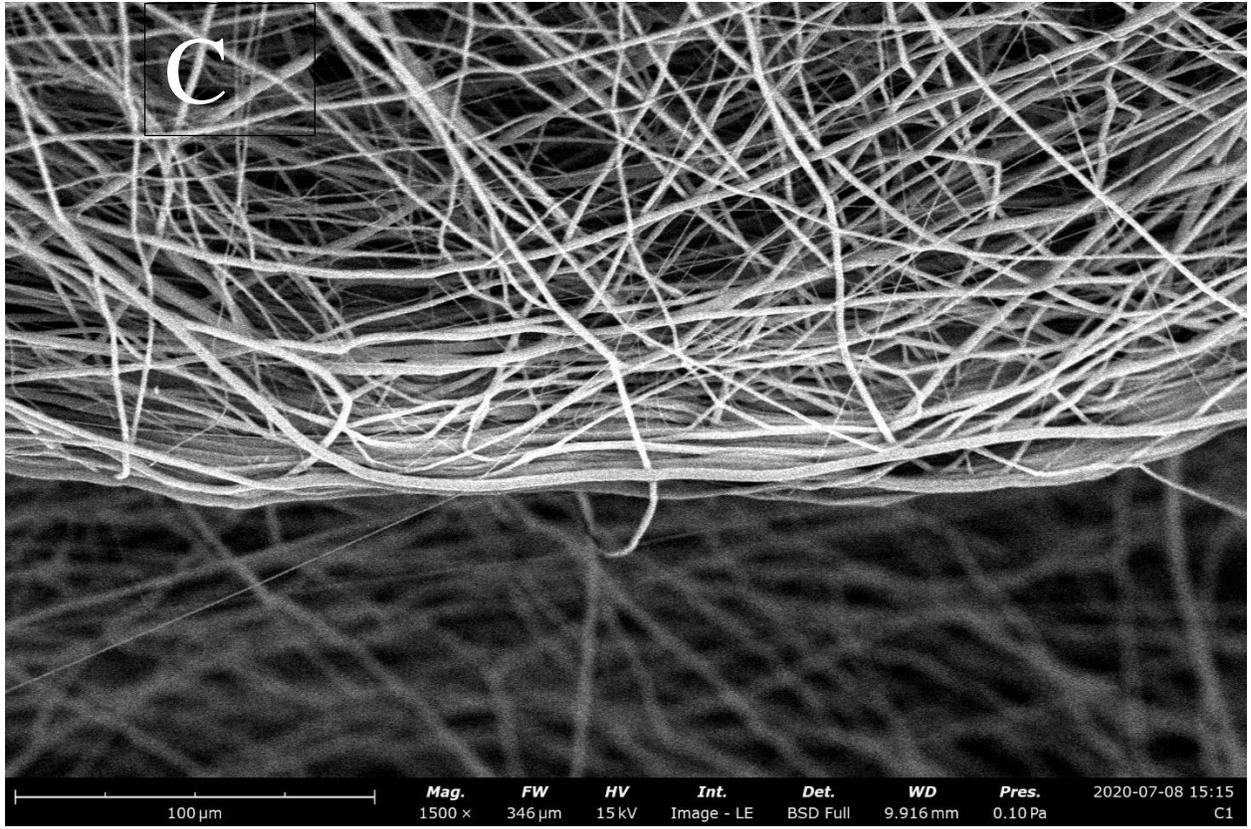

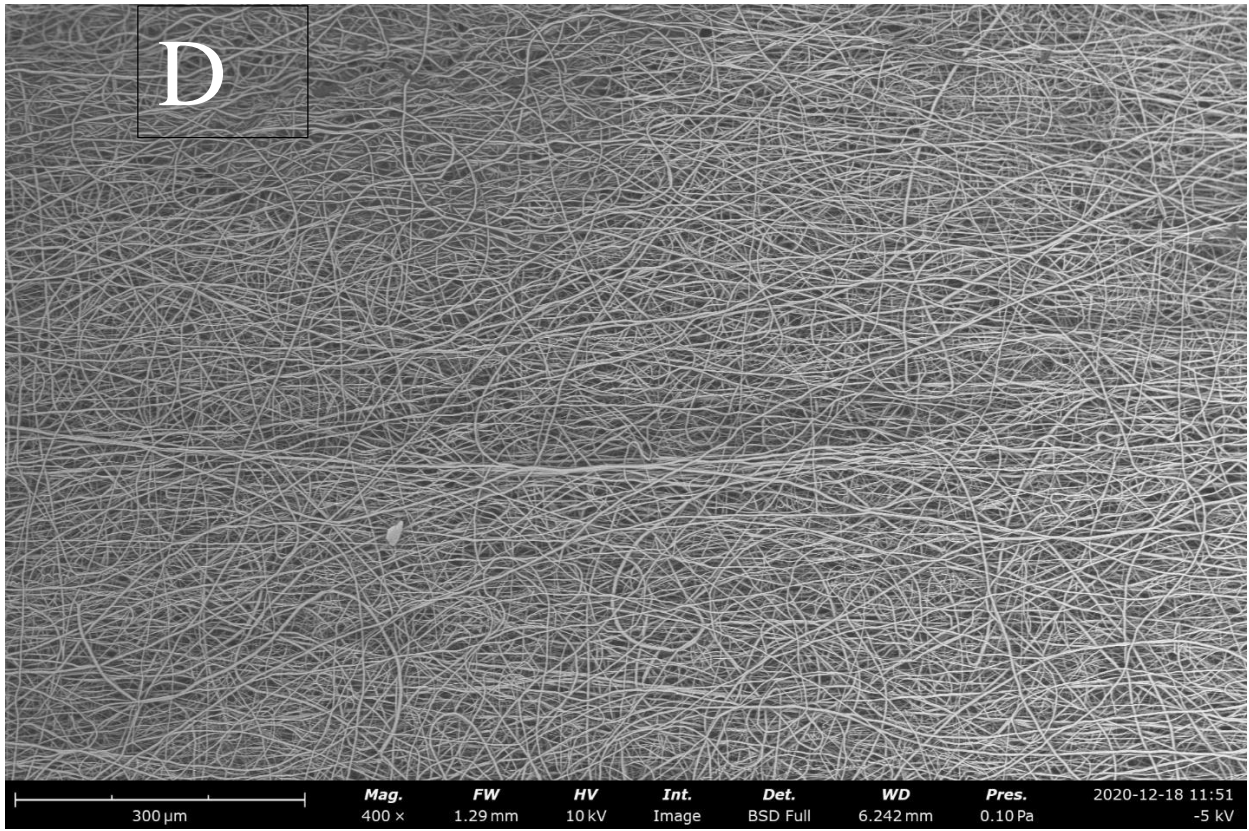

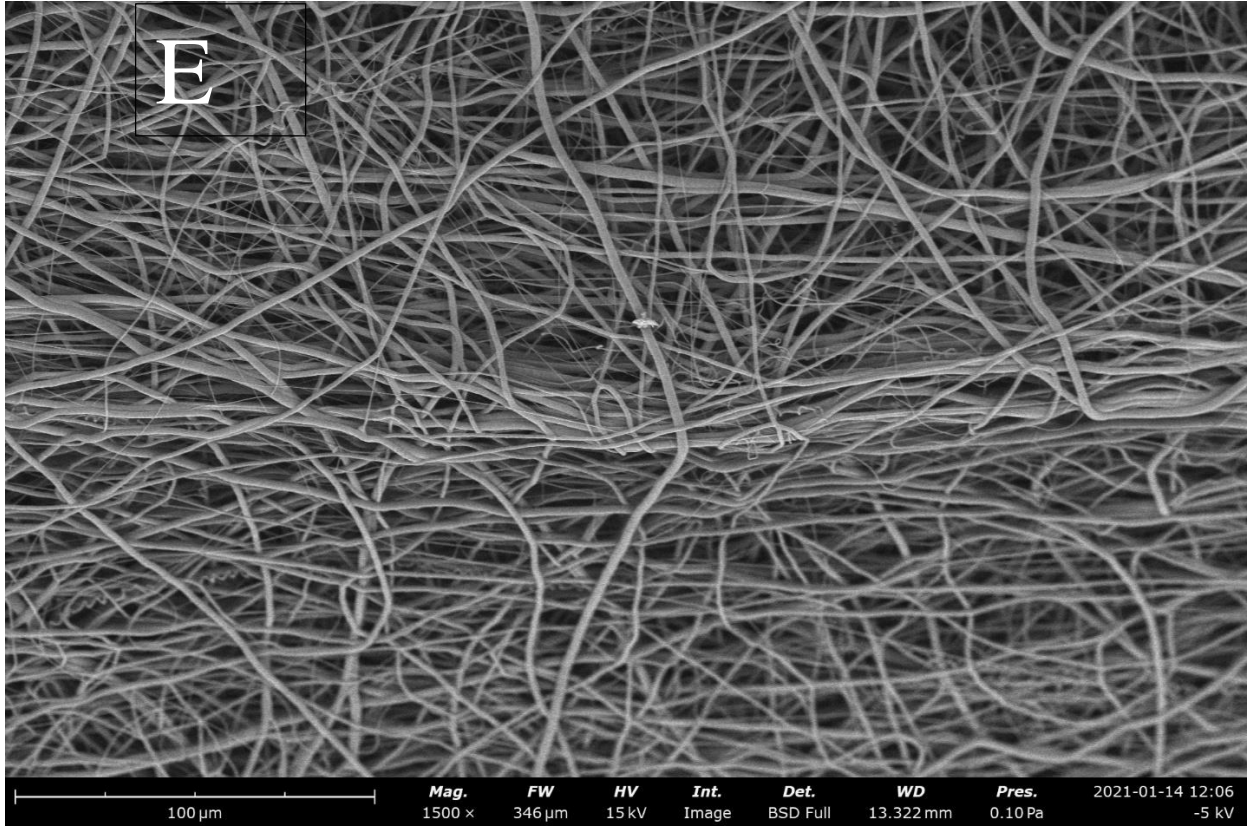

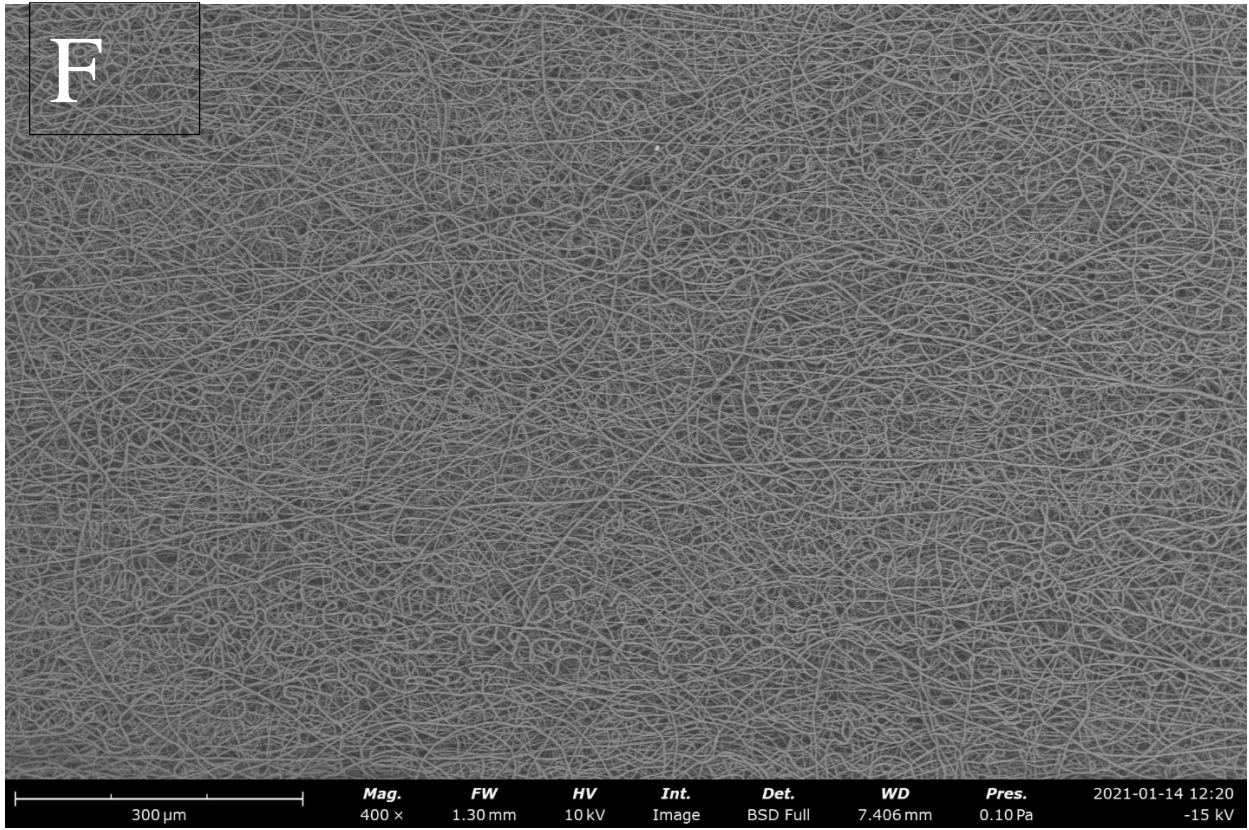

Figure 4. SEMs of fiber spun at 0 (A, B, C), -5 (D, E), and -15 (F) kV bias values. The mandrel axis runs horizontally in these images; the small regions of oriented fiber (A, B, C, D, E) also run parallel to this axis. The close-up images (C, E) are taken at a 45° angle of tilt. (A) is stitched from 50 individual SEMs to better show the spatial abundance of domains and 'bundles.'

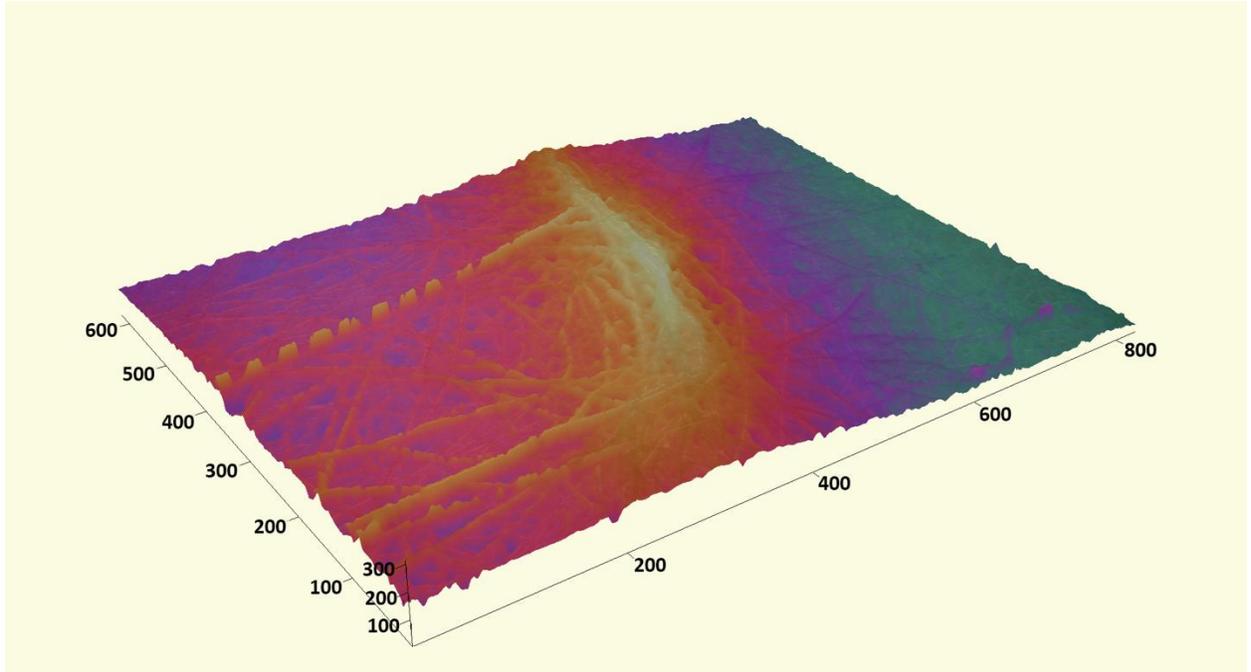
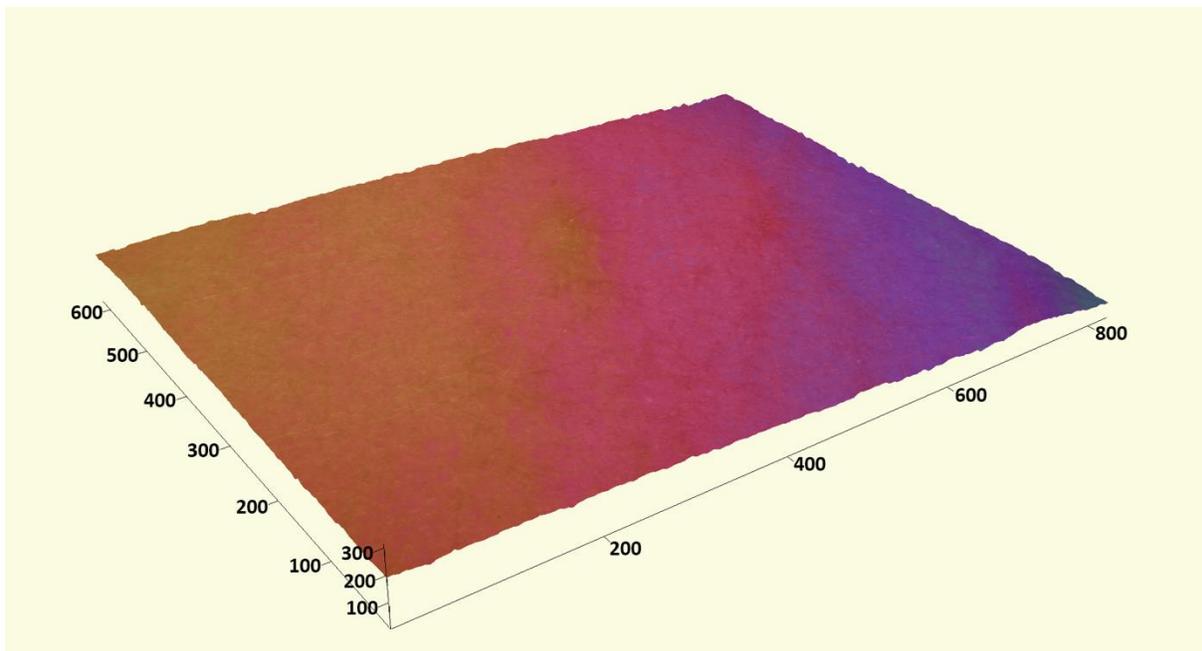

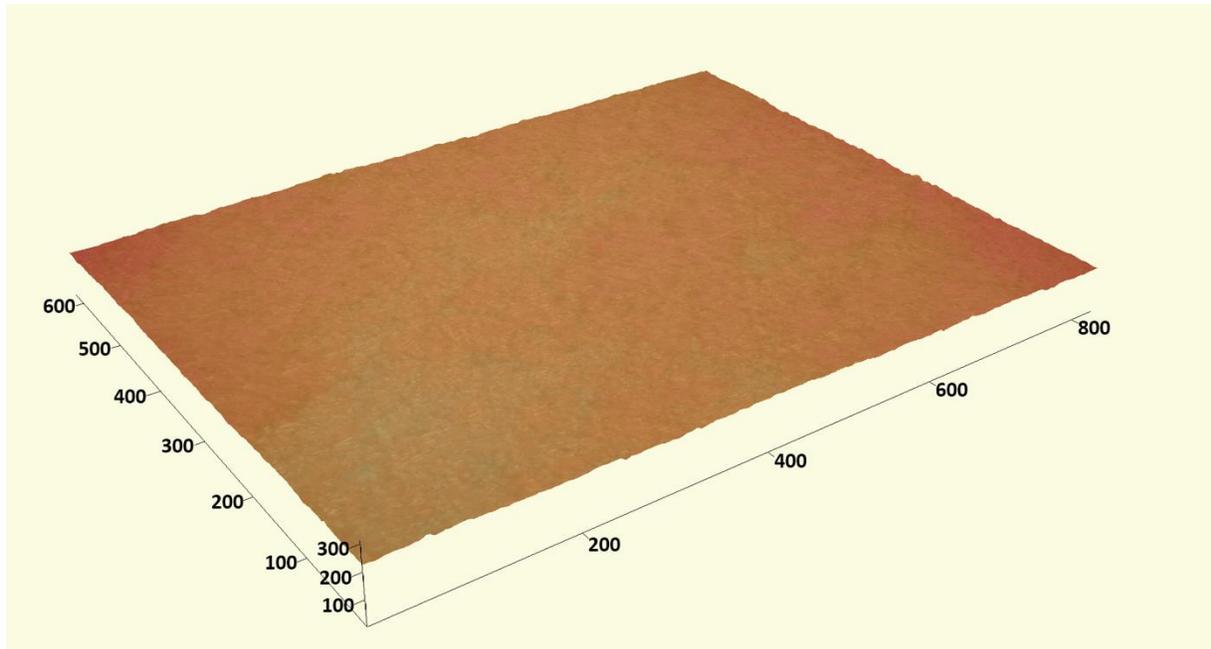

Figure 5. Optical profilometry of features found on the surface of the 0 (A), -5 (B) and -15 (C) kV depositions. Each image covers a ~0.83 x 0.62 mm area of the overall surface. The 0 and -5 kV depositions show evidence of spindle formation, albeit at different scales.

As a more compact description of orientation (versus the histograms typically presented [35, 36]), Figure 6 shows violin plots of orientation taken from the stitched images (Figures 4A, S2B, and S2D) that each consist of 50 individual SEMs as described in the *Materials and Methods* section. All three conditions show that the predominant fiber orientation is along the axial direction (0°). As attraction increases, this orientation becomes even more predominant, evidenced by the decreased spread of the violin plot as the voltage decreases from 0 to -5 and -15 kV. Between -5 and -15 kV, no significant changes in orientation are observed. Histogram representations of this data and SEM images colorized based on orientation can be found in Figures S2A, S2C, S2E, and S3.

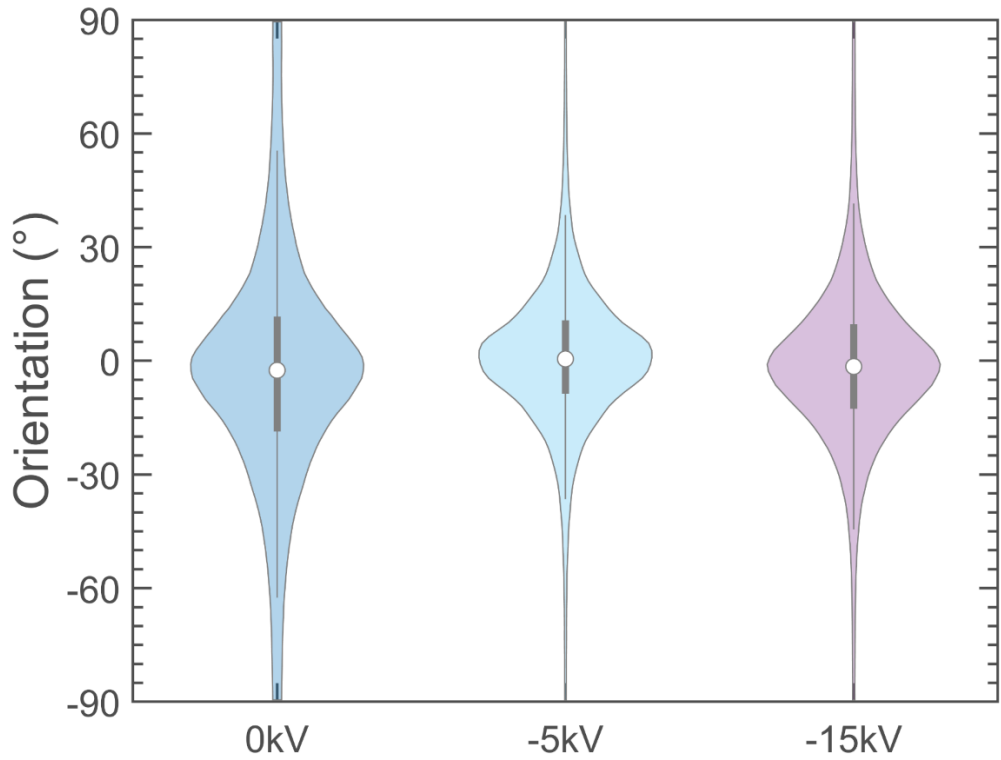

Figure 6. Violin plot depicting the distribution of fiber orientations following the 0, -5, and -15 kV depositions. The median, interquartile, and 0th to 100th percentiles are plotted as circles, short/bold vertical lines, and simple vertical lines, respectively. The 0° orientation represents the axial direction; ±90 degrees represent the azimuthal (rotational) direction of the mandrel. The corresponding SEMs and histograms are available in Figures 4A, S1, and S2.

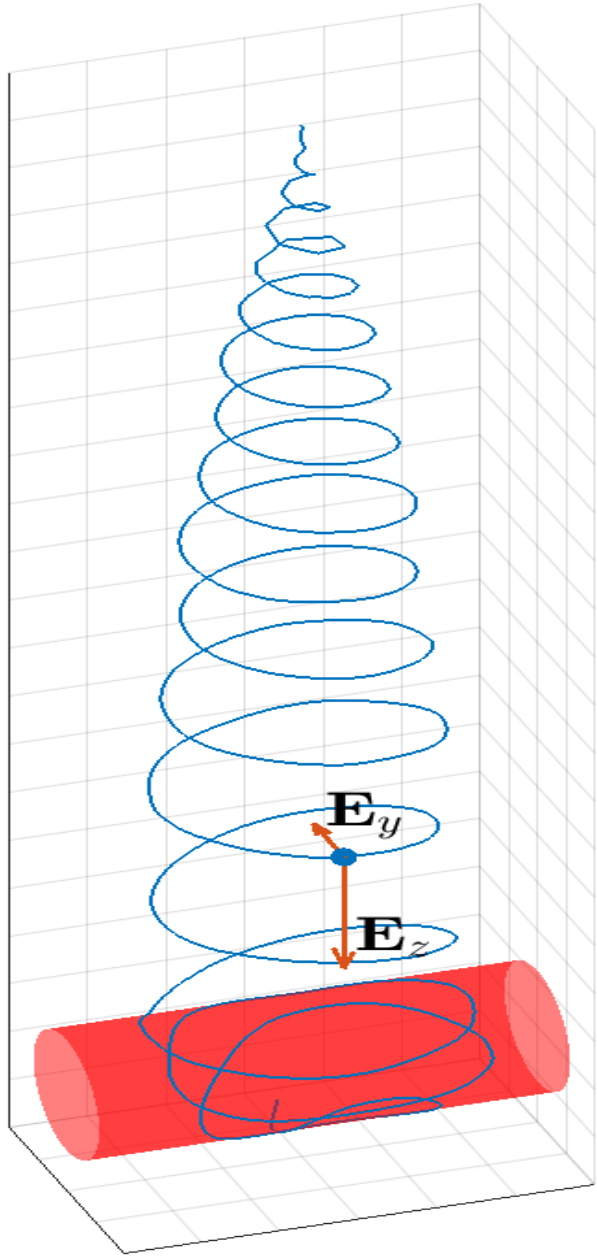

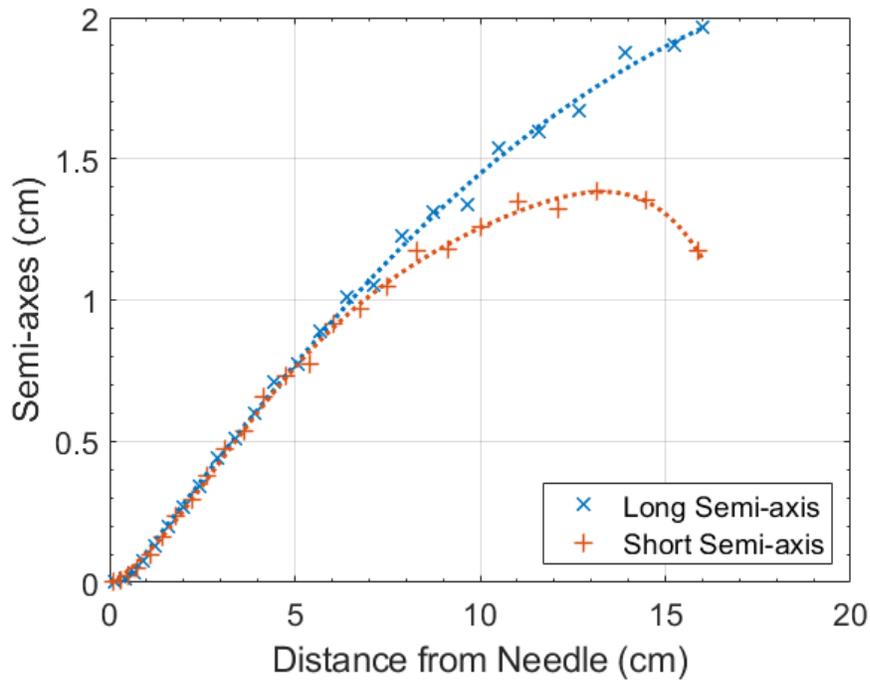

Figure 7. (A) Simulation results (animation available on-line) showing electrospun fiber approach toward a cylindrical, rotating mandrel having a non-uniform electric field represented by $\mathbf{E}_y$ and $\mathbf{E}_z$. (B) The presence of $\mathbf{E}_y$ suppressed the development of the envelope cone along its direction, resulting in a preferred final fiber orientation along the collector axis.

We believe this axial alignment could be attributed to the specific electric fields incurred during the experiments that should always point toward the long axis of the mandrel. To help illustrate this, the system of equations ((3) through (6)) were evaluated numerically, with bead insertion [27] and adaptive refinement [26] procedures proposed by Lauricella et al. Simulation parameters adapted from [27] are summarized in Table S1. As shown in Figure 7A, the horizontal cross-section of the bending cone evolved from a classically circular [2] to an elliptical shape under the influence of the horizontal component of the electric field strength $\mathbf{E}$. Its long axis along the $\hat{x}$ direction starts out almost identical (in length) to the short axis along the $\hat{y}$ direction, but grows ~50% larger than the latter before deposition (Figure 7B), where $\mathbf{E}$ is strongest. This elliptical geometry is then preserved in the deposited jet segments, resulting in a preferential orientation aligned with the mandrel axis. Circular deposition patterns have previously been observed [37] but without the preferential orientation observed here and rationalized by our non-uniform electric field strength prediction.

In Figure 8, a three-dimensional depiction of porosity variations for three samples representing each bias is shown. Again, no significant correlation between azimuthal coordinates and porosity was observed, so the remainder of the samples are plotted along the axial direction in the same manner as Figure 3, where the origin represents the peak in deposit thickness, and outliers identified by the three scale MAD criterion removed [34]. In Figure 9, the 3D data (Figure 8) for all samples is averaged better capture the data from all samples in an easily viewable format. At 0 kV, a concave-down central region of higher (~93-94%) porosity exists bounded on either side by reasonably symmetric, parabolic decreases in porosity to ~87-89%. Figure 3 clarifies that these parabolic decreases in porosity occur in regions that are 100-200 microns in thickness and constituting only ~15% of the overall volume of each sample.

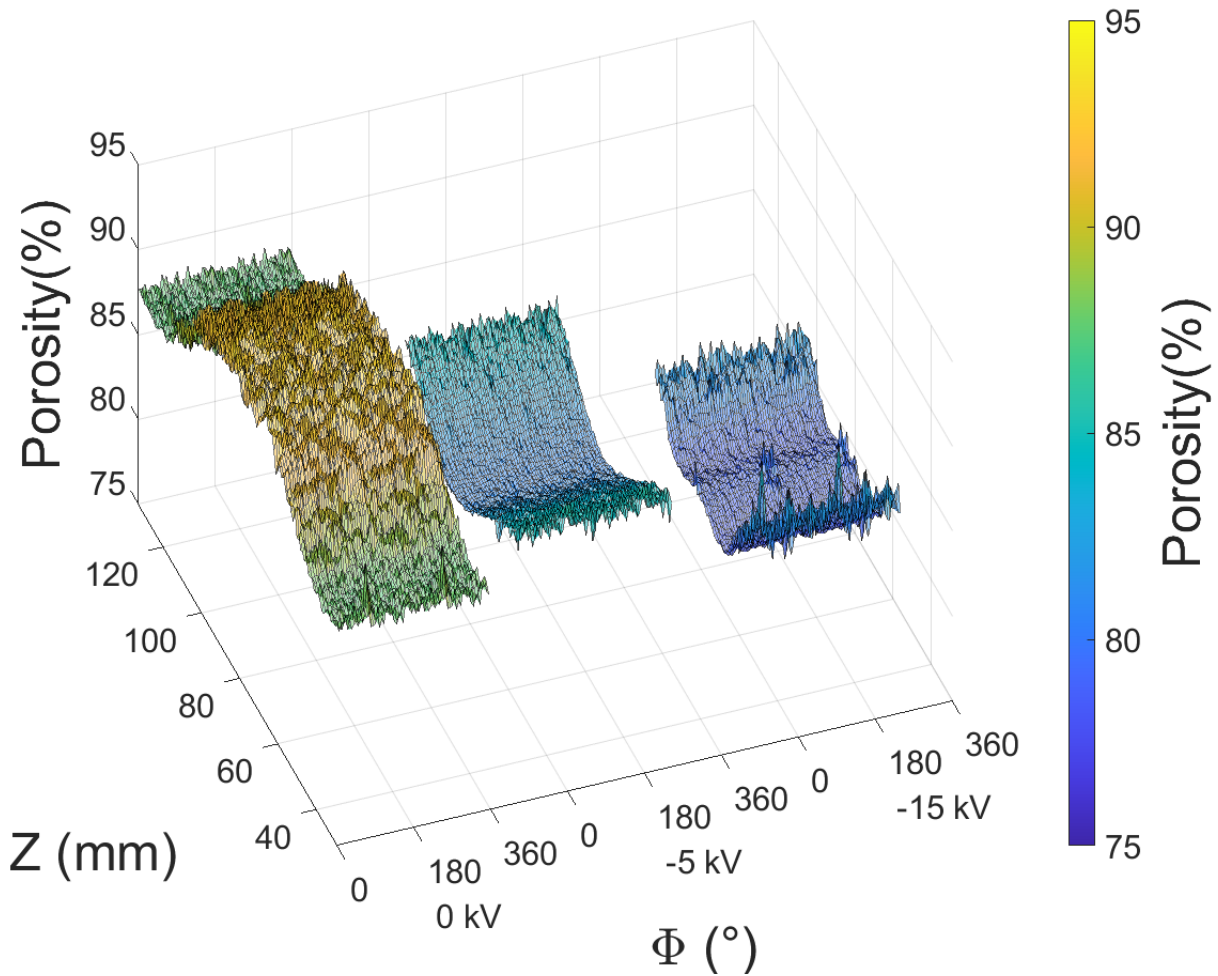

Figure 8. Porosity maps of fiber deposition for depositions on mandrels biased at (A) 0, (B) -5, and (C) -15 kV.

At -5 kV, Figures 8 and 9 reveal different characters of porosity than that following deposition at 0 kV. A concave-*up* behavior is now present in which the central porosity is ~83-84% bounded by highly symmetric, parabolic *increases* in porosity to ~85-86%. At -15 kV, the character of the porosity profile deviates from those of 0 and -5 kV. Figure 9 shows a mixed character, including both concave-up as well as a linear decreasing porosity.

Notably, the character of these porosity variations is neither visually apparent (Figure 2) nor discernable in the thickness profile (Figures 1 and 3) in which all three bias conditions show gaussian thickness behaviors. The 0 kV porosity values are the highest (at maximums of 93-94%) centered at the peak value of thickness surrounded by substantial decreases toward the edges. In contrast, larger negative bias values result in consistent decreases in porosity (typically below 85%) accompanied by small increases in porosity toward the deposition edges. In addition, the novel features observed via SEM and OP are clearly associated with the central 0 kV regions having the highest porosities.

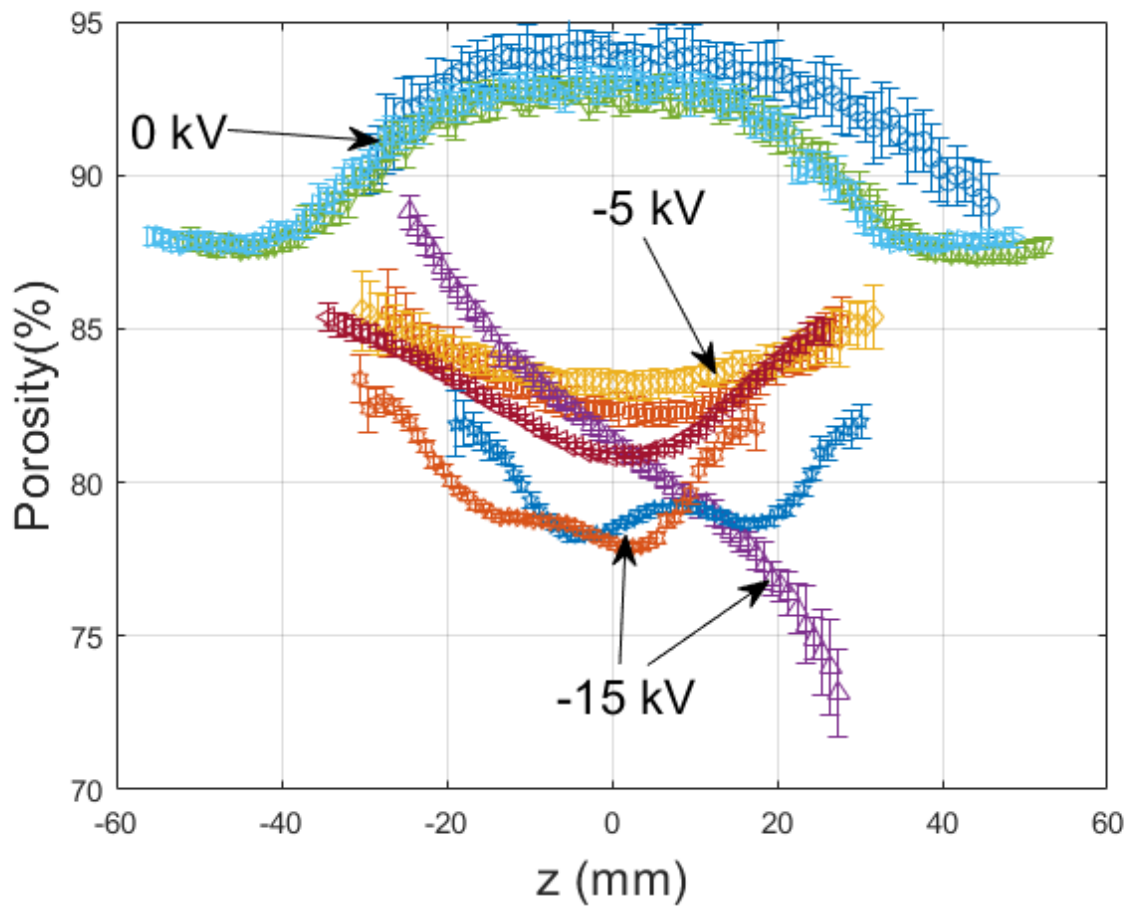

Figure 9. Two-dimensional porosity variation associated with three fiber depositions onto mandrels biased at 0, -5 and -15 kV (3 for each group for a total of 9). Each datapoint and error bar set represents the mean and one standard deviation, respectively, of 72 porosity values (less outliers) at different rotations for the corresponding axial position (z). Each bias condition falls within a distinctive group.

**Discussion**

Electrospinning is a voltage-driven process in which electrohydrodynamic phenomena at the modified Taylor Cone drive the formation of fibers from either a polymer solution or a polymer melt. The charged liquid jets so formed are accelerated towards a collector with speeds controlled by the potential existing between the modified Taylor Cone and the deposition surface. While a wide variety of biases have been explored [11-15], it is well understood that the yield of electrospinning increases with bias. Higher values of bias logically decrease 'drifting' fiber that may not otherwise deposit on the desired surface. This can, in turn, create a significant driving force towards higher voltages if expensive biological or polymeric compounds are being spun [38-43].

It is well-established that a bias on the collector surface opposite in polarity to that on the modified Taylor Cone improves fiber deposition efficiency [20] or can be used to achieve specific deposition patterns [44]. A non-zero bias on both the needle and the collector surface involving the **same** polarity, on the other hand, obviously decreases deposition efficiency but can provide **higher** values of porosity [21] that could also be valuable in specific applications.

By utilizing our prior expertise in laser micrometry [45-48] and servomechanisms [48], we proved that accurate, non-contact measurements of porosity were possible [21]. Variations in both humidity and dielectric constant showed strong influences on porosity (ranging from 0 to 90+%) that could be achieved based on deliberate variations of these two factors. This enables a unique appreciation of electrospun thickness and porosity variations along the axial direction of a cylindrical deposition versus spinning conditions; these localized changes logically affect biological integration [49]. In this work, we use the same underlying principle but expand this capability into the azimuthal direction, enabling both 2D and 3D 'mapping' of thickness and porosity distributions.

Thickness profiles plotted against both axial (Z, in mm) and azimuthal ($\Phi$, degrees (°) of rotation) create a cylindrical projection of the electrospun surface that allows for 'unrolling' and, thus, a straightforward illustration of the data. The most noticeable feature in the thickness profiles is the abundant surface roughness present following deposition at 0 kV. This contrasts sharply with the relative lack of such features following deposition at -5 and -15 kV. This distinction is obvious even at the visual level. Table 1 lists the $R_a$ values summarizing the surface morphology of the entire (~200 x 9.5$\pi$ mm) cylindrical deposit for each bias condition. Statistical analysis shows that the impact of bias on $R_a$ is highly significant ($p < 0.001$). The 0 kV bias produces $R_a$ values (9.8 µm) significantly ($p<0.001$) larger than either -5 or -15 kV. No statistical difference ($p > 0.05$) in $R_a$ was observed between the -5 kV (1.4 µm) and -15 kV biases (1.3 µm). With $R_a$ values at least 7 times greater than its higher bias counterparts, the character of the 0 kV deposition clearly pointed to the existence of a previously unreported phenomenon.

Further investigations using OP were able to characterize the features responsible for this roughness. OP provides a useful bridge between the laser-derived data and the SEM images; too often, SEM images are chosen to represent an entire spun surface. OP enables a mesoscopic examination of the surface that also provides quantitative height data, data which cannot be reliably/easily determined from SEM. The 'bumps' visible in Figures 1 and 2 were found to consist of ridged structures that sit ~210 and ~30 µm above the surrounding surface for the 0 and -5 kV depositions, respectively.

OP also shows that fibers adjacent to these ridged structures appear to be radiating away from its location. The roughness observations are clearly complemented by OP, which pinpointed the center of the specific domain microstructures subsequently revealed by SEM to contain 'bundles' of tightly aligned fiber oriented parallel to the mandrel axis. SEM was then used to examine these features further. Each ridge consists of an intriguing feature: a central region consisting of a bundle of highly oriented fibers. In tilted view, SEM imaging of one of the 0 kV 'bumps' revealed that it consists of many axially aligned fibers tightly stacked in a vertical configuration. Much smaller arrays of fibers are seen in the -5 kV deposition. These features are absent from the -15 kV deposition.

In prior work, similar surface features also appeared in the linear profiles [21] after spinning at relatively high (50% RH) levels of humidity. This work differs in that relatively low (30% RH) humidity levels were used, decreasing the potential effects of atmospheric moisture as a factor. In addition, in the prior work, this precise nanofiber alignment was not observed; in contrast, the entanglements/'conglutinations' [50] present displayed no net alignment [21].

The observation of 'curly' fibers in the SEMs of the -5 and -15 kV bias depositions indicates buckling instabilities [37, 50], a phenomenon observed during the impingement of low viscosity electrospun jets on a hard collector surface. This observation combines neatly with prior observations regarding the influence of residual solvent on porosity [21]: as the fiber moves toward the mandrel more quickly due to enhanced attraction, (1) its residual solvent content upon arrival is more likely to be preserved, leading to a lower viscosity [37] less resistant to buckling instabilities and (2) higher velocities upon collision with the collector are more likely to cause both decreased porosity [21]/"denser packing" [15, 37, 49] and increased buckling. In the absence of a rapid collision between the falling fiber and the collector, buckling instabilities are not present in the SEMs describing the microstructure of the 0 kV bias deposition.

Even though values similar or identical to the 10 kV applied needle bias in combination with a grounded (0 kV) collector are widely prevalent, no reports of such 'bundles' of highly aligned fiber exist in the extensive literature associated with the electrospinning of PCL. Via the stitching and colorization of ~50 SEMs, they could be easily detected and tracked as a function of bias. The presence of the specific surface domains, whose microstructure and visual appearance was well-demonstrated in Figures 2, 4, and 5, is also observable by laser micrometry as Figure 1 clearly shows the roughness varies with axial position. This same feature is mainly eliminated in the -5 kV bias groups and could not be detected following the -15 kV deposition. Only a few occasional surface domains/'bumps' are observed in the OP of the -5 kV deposition (Figure 5B); these are rarely observed in the laser micrometry data.

As was done previously [21], we compared as-deposited and densified thicknesses using equation (1) to develop 'maps' of porosity distribution (Figure 8). Although 3D porosity maps are only shown for one specimen of each bias, all the data is averaged and plotted against the axial direction in Figure 9. The maximum thickness following deposition at 0 kV tracks with the maximum porosity. In contrast, the maximum thickness for the -5 kV deposition is associated with minimum values of porosity. The total average porosity for each specific condition is reported in Table 1, showing that the total average does track with the values expected upon observing the 3D 'maps.' Although it was anticipated that porosity might decrease as collector bias increased, these changes in the average porosity were accompanied by dramatically differences in the shape of the porosity distribution. For 0 and -5 kV, the porosities show symmetric, parabolic behavior in which the vertices co-located with their deposition peak (at

z=0) but display opposite concavities. On the other hand, the -15 kV bias produces a mix of different profiles potentially suggesting extreme sensitivity to small changes in process conditions.

The other features of the thickness profiles in Figure 1 are as expected. Thickness distributions are all gaussian showing a peak approximately beneath the position of the 10 kV charged needle. The influence of bias on thickness is also evident in that peak thickness values (the mean of the maximum thickness for each rotational angle) are almost halved from 0 (0.76 mm average) to -5 (0.43 mm) and -15 kV (0.44 mm).  The latter two values are not significantly different ($p > 0.05$) from each other while their differences versus the 0 kV values are distinctively significant ($p<0.001$). FWHM decreases nearly linearly from 0 kV (56.3 mm average) to the -5 (46.9 mm) and -15 kV (37.0 mm) bias values.

It should not be surprising that a large net bias can suppress as-deposited porosity.  Such a bias enhances the attraction and speed of falling fibers as they move to the collector, causing them to pack into a denser deposition.  As was pointed out previously [21], the presence of residual solvent exacerbates these trends.  Perhaps more interesting is that porosity uniformity also varies substantially with collector bias. Versus a grounded (0 kV) mandrel, porosity decreases by ~6% from the center to the edge; with a -5 kV applied bias, it increases from center to edge by ~4%.

The substantial decreases - ~45% - in peak thickness observed as the bias increases from 0 kV to -5 and -15 kV, reflects the increasing strength of the net potential attracting falling fiber to any available surface of the exposed mandrel.   Unappreciated by many is that deposition acts as an effective insulation influencing the behavior of arriving fiber jets, altering the local electrostatic fields they experience and ultimately the deposition microstructure. To this end, slight variations in microstructure across the thickness are expected and a constantly increasing collector bias may be required to mitigate this effect.

A negative bias is usually applied on electrospinning collectors [51-53] to improve efficiency, yet this may, as we demonstrate, have undesired effects on electrospun microstructure, particularly in inhibiting cell infiltration [51]. As porosity decreases at -5 kV and decreases even further at -15 kV, cell penetration becomes more difficult without engineered porosity [5]. Resorption of these lower porosity areas would also likely be slower as fluid transport is hindered. Alternatively, the 0 kV deposition shows porosity values > 95%, suggesting that cell infiltration could be considerably more efficient even in the presence of the domains containing highly aligned – and obviously low porosity – fiber at their centers that would likely resist penetration but only locally at the level of these features.

The presence of fiber oriented in the axial direction (Figure 6) was unexpected. Alignment during deposition onto cylindrical surfaces has long been achieved by employing relatively high linear speeds [35, 36] but at 200 rpm, our experiments utilize linear speeds well short of those values. On the other hand, such alignment could be used to assist in the cellular motion along the axial direction of the cylinder due to the well-known effects of fiber orientation on cell migration [54-56].

Since the introduction of a modern practical and theoretical framework for electrospinning by Reneker et al. [2, 25, 50, 57], considerable progress has been made in modeling individual jet behavior. However, quantitative knowledge predicting electrospun deposit porosity is lacking; such calculations have, however, been carried out for a kindred process, solution blowing [25, 58, 59]. Given the unexpected discovery of alignment in the axial direction of the mandrel, we sought to model jet behavior to attempt to understand this phenomenon. In most, if not all previously published work, the external electric field strength **E** was approximated as $-(\varphi_0/h)\hat{\mathbf{z}}$, a term describing a uniform electric field pointing downward to an infinite plate collector (the *xOy* plane) from a positively charged electrode. However, to better model the electric field surrounding the cylindrical mandrel in our experiment, we considered an infinitely long, uniformly charged cylinder having a radius of $a_c$ around the x-axis. The electrical field strength around such a cylinder is proportional to the reciprocal distance from its axis *1/r* and a pre-factor consisting of surface charge density $\sigma_e$, air permittivity $\varepsilon$, and $a_c$ [60]:

$$\mathbf{E} = \frac{4\pi \sigma_e a_c}{\varepsilon} \frac{1}{r} \mathbf{e}_r \qquad (7)$$

where $\mathbf{e}_r$ is the unit vector along the positive radial direction. Since $\sigma_e$ was not measured in the experiment, we derive the pre-factor $4\pi\sigma_e a_c/\varepsilon$ from relation $\mathbf{E} = -\nabla\varphi$ along with the following boundary conditions: (1) the reference potential on mandrel surface, $\varphi(r = a_c) = 0$ and (2) the potential $\varphi_0$ occurs when the distance *r* to the cylinder axis equals *h*, $\varphi(r = h) = \varphi_0$. Adapting to Cartesian coordinates, we reach an expression for **E** using a few clearly defined, easily controlled input variables in equation *(6)*. The negative sign in this equation emerges from the direction of potential gradient and effectively indicates the collector cylinder is a cathode. As Figure 7 makes clear, during simulation the horizontal component of **E** has led to a preferred orientation along the collector's long axis in both the traveling and deposited jet.

While helpful in explaining the unexpected axial orientation, some limitations in the modeling shown in Figure 7 exist. The effects of gravity and aerodynamics are not considered as their effects are widely considered to be negligible [2]. The relatively slow linear speed of the 200 rpm mandrel rotation (200*9.5*π/60 ≈ 99 mm/s) in the experiment should not lead to fundamental changes in arriving jet behavior. In addition, we know that an instantaneous electric discharge is associated with each node following deposition, so the potential influence of accumulated laydown on the arriving jets is overlooked. This step ensured that the simulation remained computationally feasible as the electrostatic term in equation (4) scales as $O(n^2)$ with the number of nodes. Consequently, the model only truly describes the behavior of the initial layer of deposition.

Other than exploiting the algorithms developed for rapid computation of n-body problems, such as the fast multipole method [61], evaluating the effects of deposited nodes could be achieved through modifying the external electric field, **E**. The deposition of electrified jets onto opposite-signed collector results in a decrease of surface charge density $\sigma_e$ and, as equation *(7)* points out, reductions in field strength. This will, in turn, lead to lower incoming fiber velocities and more gently formed, more porous [62] deposition. Meanwhile, charge dissipation across the collector surface should slowly decay this "cushioning" effect, and the complex dynamics of charge delivery versus dissipation will likely determine the laydown porosity. For example, a thick pre-existing laydown could slow down charge dissipation hence producing higher porosities while the presence of humidity achieves the opposite effect via charge "bleed-off" triggered by the presence of ambient water molecules. The influence of applied bias is more complicated as it has interplay with both incoming charge density and discharge rate following deposition. However, as Figures 7 and 8 suggests, in this study the net effect of increased bias was to decrease porosity.

Considering the spatial distribution of this process of charge delivery and dissipation could enable more detailed modeling of electrospun deposition. Models correlating the incoming velocity and diameter to thickness and porosity of small areas of deposition have already appeared in the literature [63, 64]; these were found to predict the rate of charge decay [65].

Discretizing the collector surface into a mesh and apply both models to each cell will result in charge distribution from which electric field **E** could be computed via integrating Poisson's equation. Substituting this **E** into equation (4) could then predict the deposition coordinates and velocity that serves as input for the porosity/thickness and charge decay model. Repeating this procedure for extended sessions will, in principle, predict the thickness and porosity distributions observed experimentally.

The use of much higher rotational speeds to collect azimuthally aligned fibers is widely practiced [35, 36, 66-68]. On the other hand, axially aligned fibers have merit in certain tissue engineering applications promoting the proliferation and migration of specific cell types [69, 70]. The use of auxiliary electrodes to achieve alignment has been extensively studied both experimentally [70-72] and theoretically [73], yet few have explicitly considered the influence of collector geometry on the resulting electric field and fiber orientation. By examining the cylindrical electric field surrounding our collector, we arrive at an intuitive explanation to the unexpected axial alignment observed in all bias groups. Equation *(6)* suggests that this alignment should scale with applied potential ($\varphi_0$) and be inversely proportional to the natural logarithm of the needle to collector distance/collector radius ratio, *In(h/$a_c$)*. Given the wide use of cylindrical collectors in electrospinning, we believe this finding could provide yet another, more easily utilized (versus specialty collector designs) tool enabling the facile, efficient tuning of alignment.

Even though our observations are specific to PCL, it is reasonable to anticipate that they also appear in other polymers spun at similar solids loadings out of HFP [21]. The features evident in Figure 4 – revealing exquisite alignment on a small scale – likely relate to collapse of relatively short segments of a single polymer jet into the same small area [74, 75]. The robust, almost perfect alignment parallel to the mandrel axis suggests that it originates from the interaction between the proposed local field non-uniformities and the depositing fiber. Thus, the higher biases initially decrease domain formation (-5 kV) and then eliminate it (-15 kV) entirely as the velocity toward the cylinder increases. Thus, fiber domains containing spindles consisting of aligned fibers are initially disrupted (Figure 5B) and then disappear from the thickness surface profiles (Figure 5C). However, the alignment data (Figure 6) clearly shows that even though the existence of domains is eliminated at -15 kV, fibers continue to follow the net aligning influence of the non-uniform electric field as the dominant direction of orientation remains parallel to the long axis of the cylinder.

We also recorded hi-speed video of the laydown process (Figure 10) to make visual observations regarding the character of the process versus bias. In the videos, a considerable portion of the arriving fiber 'overshoots' the mandrel – a phenomenon especially visible at 0 kV - and then deposits on it by reversing direction and going back upward toward the mandrel. This suggests more possibilities for interaction with the non-uniform electric field [74-76] that could further influence local-scale microstructure. Higher values of bias also induce a perceivable increase in the velocity of the fiber arriving at the mandrel. In addition, some evidence for the insulating effect of as-deposited fiber exists as arriving jets can be observed to track back and forth across the mandrel surface.

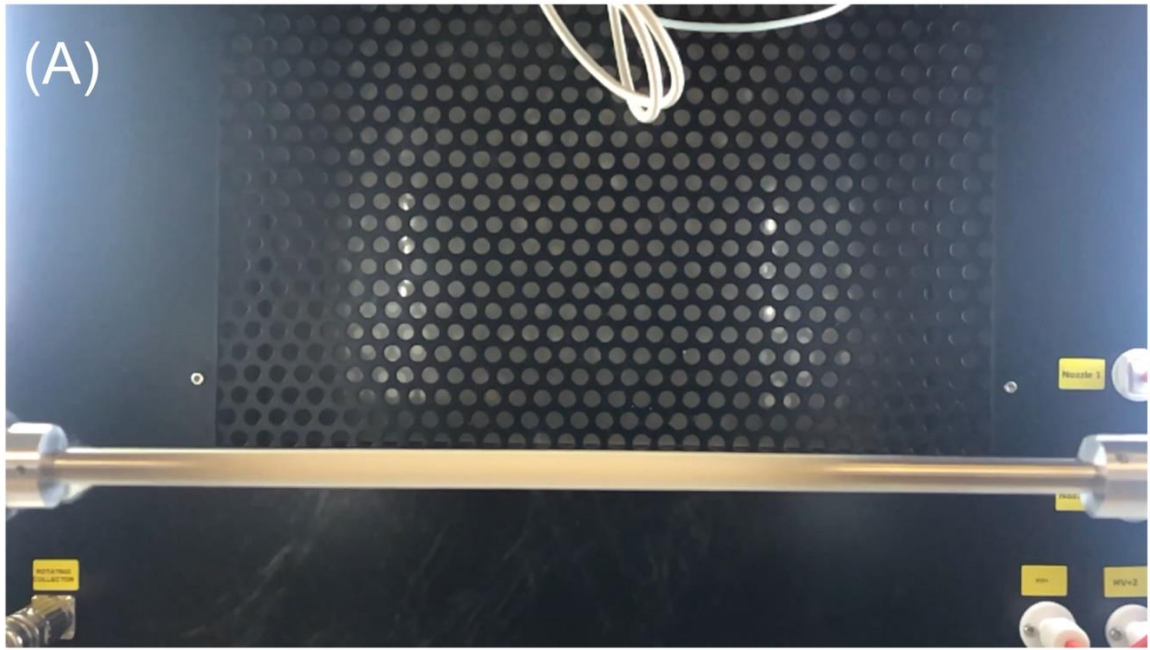
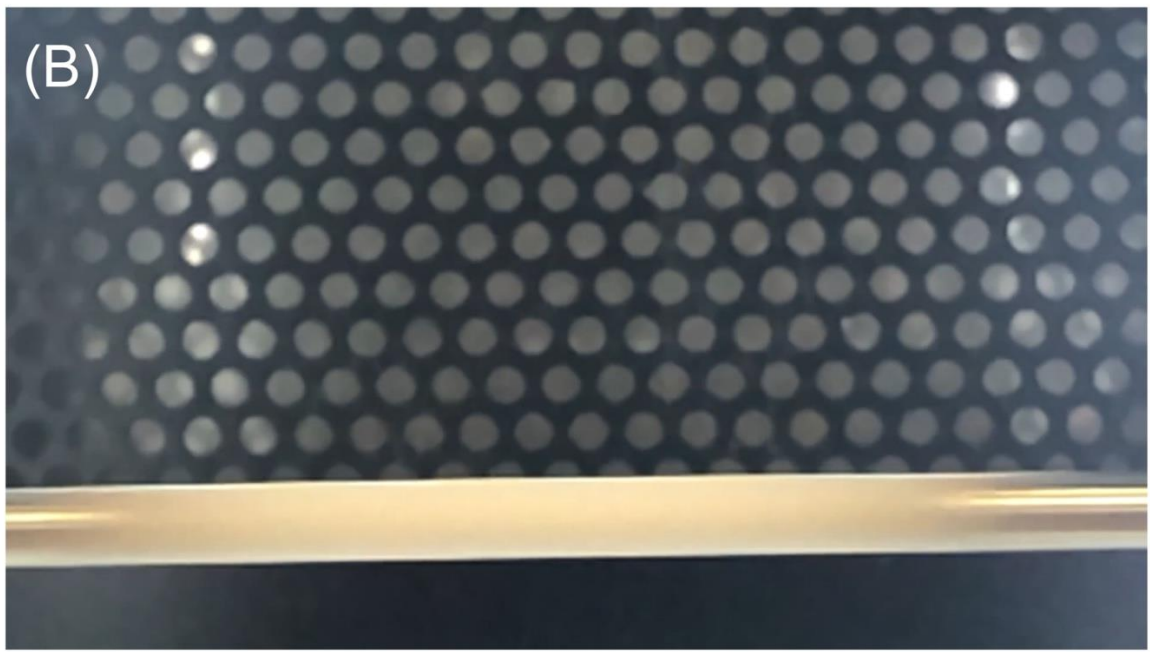

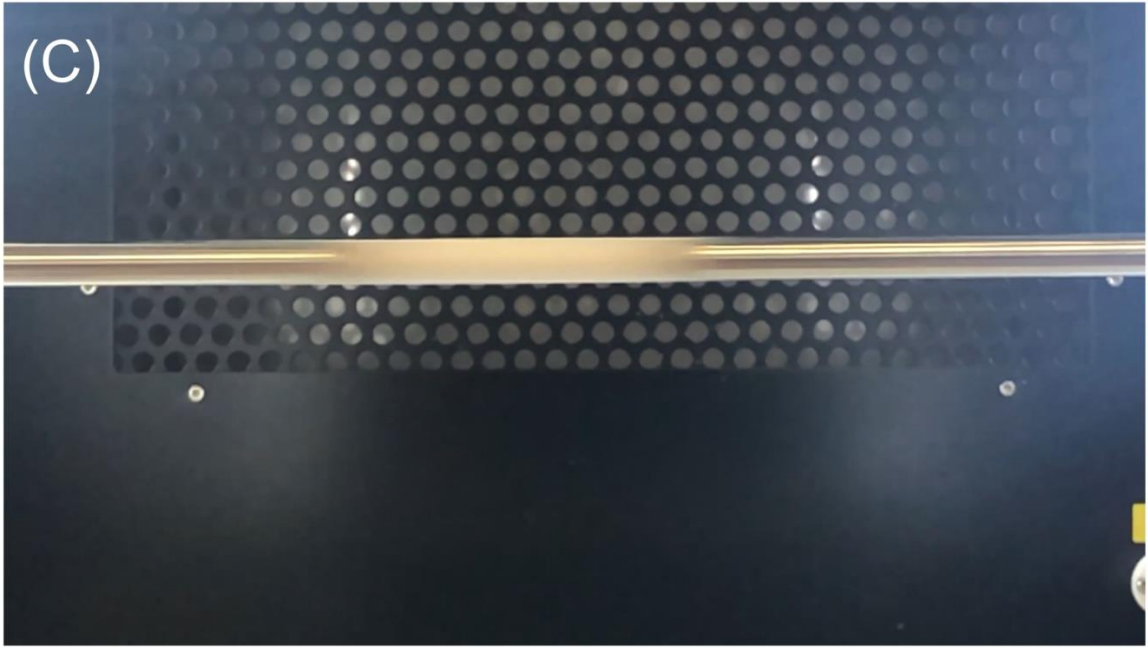

Figure 10. Still images of high-speed video (available on-line) of arriving jets at 0 (A) , -5 (B) and -15 kV (C) bias levels. Some fiber clearly goes below the mandrel at each bias and is then drawn back upwards where it achieves deposition.

**Conclusions**

Via careful manipulation of non-contact laser micrometry data, thickness and porosity profiles were created following deposition onto a cylindrical mandrel. 'Maps' revealing substantial variations in thickness and porosity versus 0, -5, and -15 kV biases were created. As net bias increases thinner, more 'focused' depositions occur. Deposition at 0 kV showed extensive mesoscale surface roughness resolved into novel, localized domains with tightly aligned fiber bundles parallel to the mandrel axis at their center. Observed trends in porosity agree with our prior observations of residual solvent effects: increased bias causes faster motion toward the mandrel, meaning that solvent content upon arrival is higher, leading to lower viscosities less resistant to buckling/compaction. In addition, higher velocities during deposition cause both decreased porosity/"denser packing" and increased buckling. Unexpectedly, we also observed substantial orientation along the mandrel axis. By modifying classical bending instability models to incorporate more realistic electric fields around a mandrel, simulation revealed that horizontal components in the modified electric field strength alter bending loop shape, causing the observed alignment.  This finding provides an easily utilized tool enabling the facile, efficient tuning of electrospun fiber alignment.

**Acknowledgments**

Research reported in this publication was supported by the Bureau of Land Management under award number L14AS00048. The content is solely the responsibility of the authors and does not necessarily represent the official views of the Bureau of Land Management. We are grateful to the Ohio Supercomputer Center [77] for providing complimentary credits enabling its computational resources to be effectively used by Ohio academic clients. The authors also thank engineers at Aerotech, Inc. and LaserLinc, Inc. for their assistance in troubleshooting the hardware used in this research.

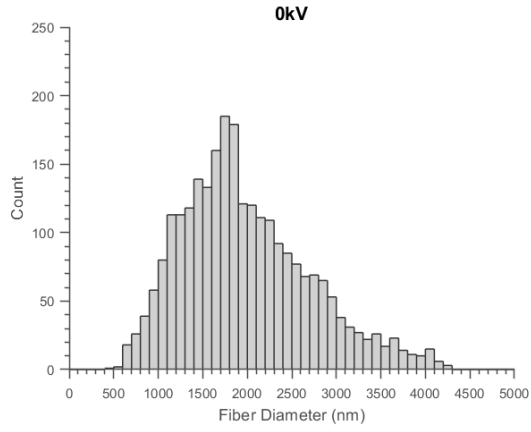
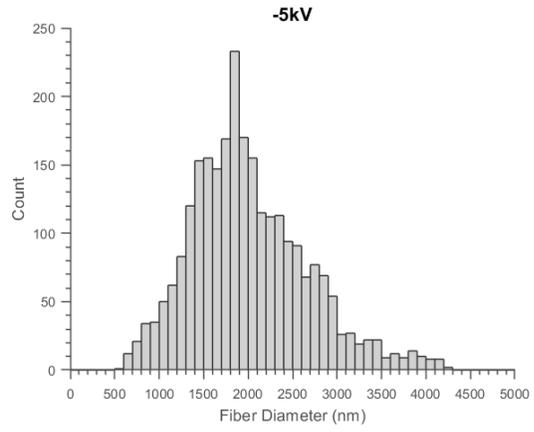
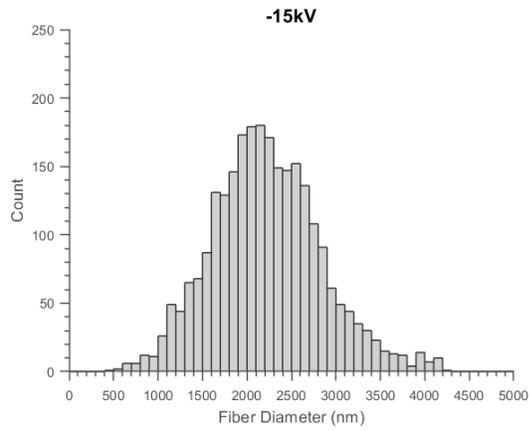

Fig S1. Distribution of fiber diameter following deposition onto the mandrels at 0, -5, -15 kV bias.

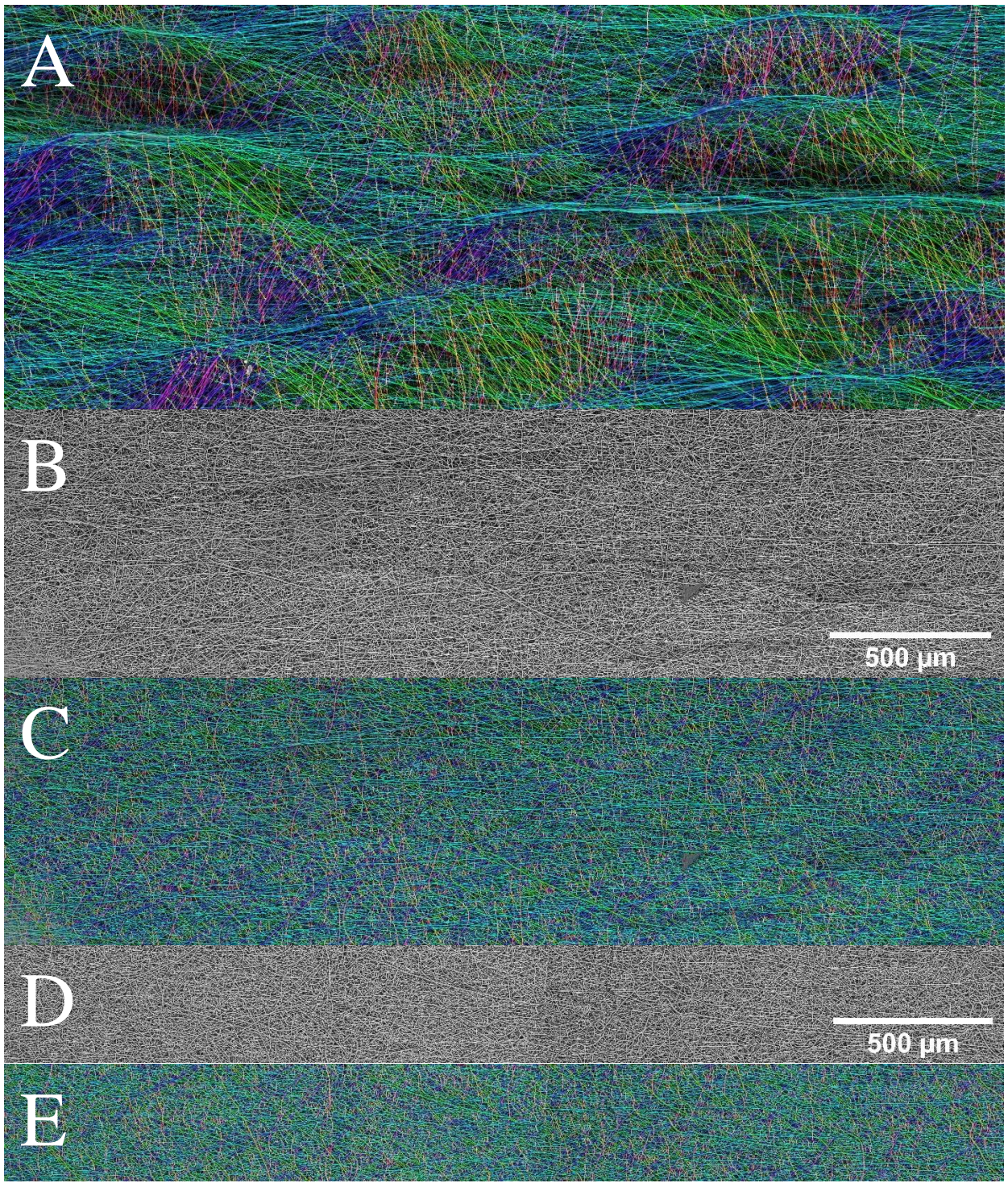

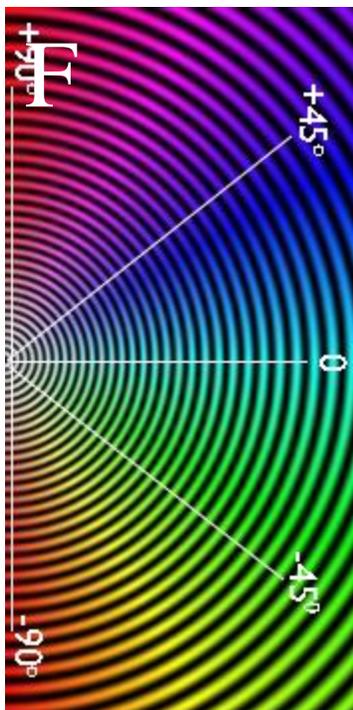

Fig S2. Stitched SEM images following laydown at (A) 0, (B,C) -5 and (D, E) -15 kV bias. The -5 (B) and -15 (D) kV depositions are normal SEM images. Orientation-colorized images are provided for the 0 (A), -5 (C), and -15 (E) kV depositions. The mandrel axis runs horizontally in these images. A colormap of orientation (F) is provided for reference.

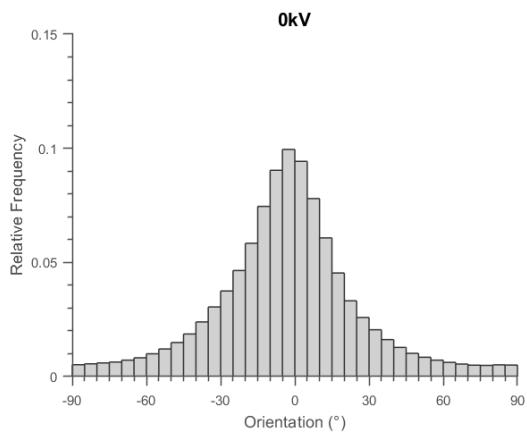
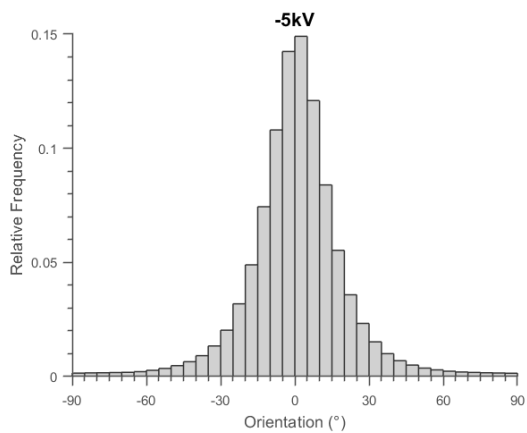

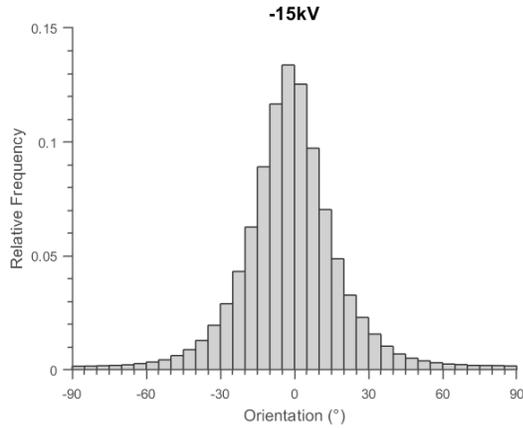

Fig S3. Distribution of fiber orientation following deposition onto mandrels at the 0, -5, -15 kV values of bias.

Table S1. Simulation parameters used in generating the results shown in Figure 9. All values are reported in cgs units. The needle-to collector distance is the distance from needle to the central axis of the collector cylinder. Numbers were adapted from Lauricella et al. [27] with the exception of the applied potential. To intensify its influence, the applied potential $\varphi_0$ is greater than that used in the experiment.

| Mass Density | $g*cm^{-3}$ | 0.83925 |
|---|---|---|
| Charge Density | $g^{1/2}*cm^{-3/2}*s^{-1}$ | 44000 |

| Elastic Modulus | $g \cdot cm^{-1} \cdot s^{-2}$ | 50000 |
|---|---|---|
| Viscosity | $g \cdot cm^{-1} \cdot s^{-1}$ | 20 |
| Surface Tension | $g \cdot s^{-2}$ | 21.133 |
| Jet segmentation length | cm | 0.02 |
| Initial Jet Radius | cm | 0.005 |
| Initial Jet Velocity | $cm \cdot s^{-1}$ | 0.28294 |
| Perturbation Frequency | $s^{-1}$ | 10000 |
| Perturbation Scale | cm | 0.001 |
| Needle-to-Collector Potential | $g^{1/2} \cdot cm^{1/2} \cdot s^{-1}$ | 250.17 |
| Needle-to-Collector distance | cm | 17 |
| Collector Radius | cm | 1 |

Table S2. Characteristics of solutions used in electrospinning.

| Viscosity (cP) | 761 |
|---|---|
| Surface tension (mN/m) | 17.68 |
| Density (g/cm3) | 1.4098 |
| Solid content (%) | 5.36 |